\documentclass[toc]{PoS}
\usepackage{amsmath}

\title{Introductory notes on holographic superconductors}

\ShortTitle{Introductory notes on holographic superconductors}

\author{\speaker{Daniele Musso}%
         \\
        Abdus Salam International Centre for Theoretical Physics (ICTP)\\
        Strada Costiera 11, I 34014 Trieste, ITALY\\
        e-mail: \email{dmusso@ictp.it}}

\abstract{The purpose of these lecture notes is to give a quick and introductory overview of \emph{holographic superconductors}.
Besides the actual description of the standard holographic superconductor, attention is paid to the motivations and the relation
with the previous, non-holographic context.}

\FullConference{ Informal lecture notes prepared for the IX Modave school of mathematical physics\\
                 Modave, September 2013\\
                 Belgium                
                 }

\begin{document}


\section*{Foreword}

In the writing of these notes, the major sources of inspiration were the classic papers 
\cite{Polchinski:1993ii}, \cite{Weinberg:1986cq} and \cite{Hartnoll:2008kx}.
Throughout the present paper some basic knowledge of $AdS$/CFT is understood and we refer to the many 
reviews on the market, particularly to \cite{Aharony:1999ti,Zaffaroni:2000vh}. 
Albeit an attempt to encompass all the relevant literature, the amount of research being lately performed on the 
subject of holographic superconductors renders the bibliography necessarily incomplete.

\section{Brief introduction}

It is fair to say that the holographic superconductor introduced in
\cite{Hartnoll:2008kx,Gubser:2008px,Hartnoll:2008vx} furnishes the paradigmatic model to describe a spontaneous 
$U(1)$ symmetry breaking in the gauge/gravity framework. The holographic approach allows
us to qualitatively and quantitatively address some
crucial questions occurring in strongly coupled systems and proves 
particularly useful in finite-temperature circumstances when we are interested in 
the real-time response. In fact, the finite-temperature and real-time context
is usually described poorly by standard methods, either perturbative or not (e.g. the lattice).
The failure of perturbation theory at strong coupling comes with no surprise. On the contrary,
the weakness of a lattice analysis could sound unexpected. Especially because
the lattice treatment is commonly suited to analyze a strongly-coupled theory.
The biggest trouble of addressing finite-temperature, real-time
physics on the lattice resides in the ambiguities of analytic continuations and
the consequent introduction of systematical errors whose absence (in other contexts)
is typically the strong point of the lattice method itself.

\section{General motivations}

\subsection{The goals, the strategy and the philosophy}

\subsubsection{Spontaneous symmetry breaking at strong coupling}

The standard paradigms with which spontaneous symmetry breaking phenomena are described
are generally based on a weakly coupled picture and on perturbation theory. 
However, there are relevant physical systems (like those featuring quantum 
phase transitions) which often call for a generalization of the symmetry breaking ideas to 
a strongly coupled context.

An important point consists in the similarities and the differences between the standard methods
(such as the Ginzburg-Landau picture) and the holographic effective approaches%
\footnote{We refer to \cite{Musso:2013rva} which contains some specific comments on this.}.
This has both a purely theoretical and a phenomenological interest.
Regarding the former, topics like the circumvention of the Coleman-Mermin theorem represents 
interesting theoretical questions. In relation to the latter, many materials manifest complicated 
many-body behaviors whose nature could probably be unveiled with a proper understanding of 
their strongly coupled dynamics. Notably, the high-$T_c$ superconductors (both in their superconducting 
and strange metal phases) belong to this category.

\subsubsection{Strategies}

The $AdS$/CFT conjecture was incubated in the physics of black holes and large $N$ limit
of non-Abelian quantum field theory, but it was eventually born in string theory. 
There it assumed a precise (even though not proven) character and formulation.
Despite its origin, gauge/gravity ideas and techniques appear to have an important
value also independently of string theory.
A new genre of effective field theory approaches seems actually suggested by $AdS$/CFT. 
These could possibly be extended to a larger context where other analogous gauge/gravity 
dualities can be legitimately conjectured and applied.

This coexistence of different souls in $AdS$/CFT has a direct effect on the methods which 
are employed in doing research. Roughly speaking, it is possible to proceed either \emph{top-down} or \emph{bottom-up}.
The former approach means that one starts from a well defined and consistent gravity 
model (say, string theory, M-theory, SUGRA,...), the latter instead refers to an effective approach
based on simple models conceived to capture particular IR phenomenological aspects%
\footnote{To this regard, concerning the holographic superconductor, it was introduced
adopting a bottom-up approach in \cite{Hartnoll:2008kx,Gubser:2008px,Hartnoll:2008vx} and then later embedded
in a UV completed framework in \cite{Denef:2009tp,Gubser:2009qm,Gauntlett:2009dn}.}.

\subsubsection{Philosophy}

A preliminary disclaimer is in order.
Needless to say, the $AdS$/CFT research, and particularly the $AdS$/Condensed Matter Theory
($AdS$/CMT) branch, has an ambition to phenomenology. Even though being born in a rather abstract stringy
environment, the conjectured gauge/gravity duality seems to offer the possibility of 
addressing real-world problems in a fascinating new way. Enthusiasm and caution are however
both badly required.

Not only out of intellectual honesty, but also on a technical level, being able to
discern the holographic model itself from the phenomenological interpretation we attempt to give it
seems always to be crucial. The holographic superconductor is no exception and offers a neat instance 
of how we intend to do physics holographically. The point will become clearer throughout these lectures,
but the general idea is that the holographic model has to be studied both with and without the phenomenological prejudice
we have in mind. Maybe such an approach is a generic feature of theoretical physics. Nevertheless, 
such general cautious attitude becomes even more important when working in a context which is based on 
conjectures relating distinct areas of physics.

The $AdS$/CMT panorama provides us with a new class of solvable toy-models.
Quoting \cite{Sachdev:2010ch}, this could be essential to gain insight on 
wider classes of systems even when the latter happen not to be directly 
solvable.

\subsubsection{Round trip: from quantum gravity to strong coupling, and back}

As a comment, notice that $AdS$/CMT (like gauge/gravity correspondence in general)
can be employed in both directions: From gravity to field theory and vice versa. In this sense,
mapping quantum gravity models to lower-dimensional quantum field theory is appealing
for a novel experimental reason: making experiments on quantum gravity in the condensed matter lab.
Clearly, this statement is based on a maybe overenthusiastic (at least so far) optimism; at any rate,
the hope is based both on the theoretical progress in $AdS$/CFT and on the technological progress 
especially in handling systems like cold-atoms. Indeed, such systems possess the crucial
feature of being particularly tunable. Their flexibility and the control we attain, possibly,
could offer us a vast and exciting playground also in relation to rather abstract questions concerning
quantum gravity, black holes, information paradox, and on.

\section{Condensed matter (\emph{without strings...})}
\label{wout}

In the following, we will be mainly concerned with high-$T_c$ superconductors, both in the condensed phase and 
in the so-called strange metal phase. A crucial concept to be introduced at once is that of 
\emph{quantum phase transition}. Indeed, the dynamics of the above-mentioned phases of high-$T_c$
superconductors is believed to be strictly connected to the quantum critical behavior occurring in 
the vicinity of a quantum phase transition. 

\subsection{Quantum phase transitions and quantum criticality}

A quantum phase transition is a phase transition at zero absolute temperature
which is driven by quantum fluctuations instead of thermal fluctuations. 
Even though, strictly speaking, a quantum phase transition occurs only at $T=0$, there exists a nearby region in the phase diagram
for $T>0$ where the dynamics of the system is strongly affected by the presence of the quantum critical point.
This region goes under the name of \emph{quantum critical region}.

Some features of the qualitative and quantitative descriptions of a continuous quantum phase transition can be thought in analogy
with the thermal continuous phase transitions and can actually be borrowed from a theoretical approach \emph{\'a la} Ginzburg-Landau.
More specifically, a system in the proximity of a quantum critical point is 
characterized by a diverging coherence length and the behavior of the observables
is described by the corresponding quantum critical exponents. 
Strictly at quantum criticality, the coherence length is infinite and the system becomes scale invariant; this is the reason why 
the critical system can be described with a conformal quantum field theory.
This is also the context where holographic tools enter into the game, especially when the critical system happens to be strongly coupled.

\begin{figure}[ht]
\centering
\includegraphics[width=110mm]{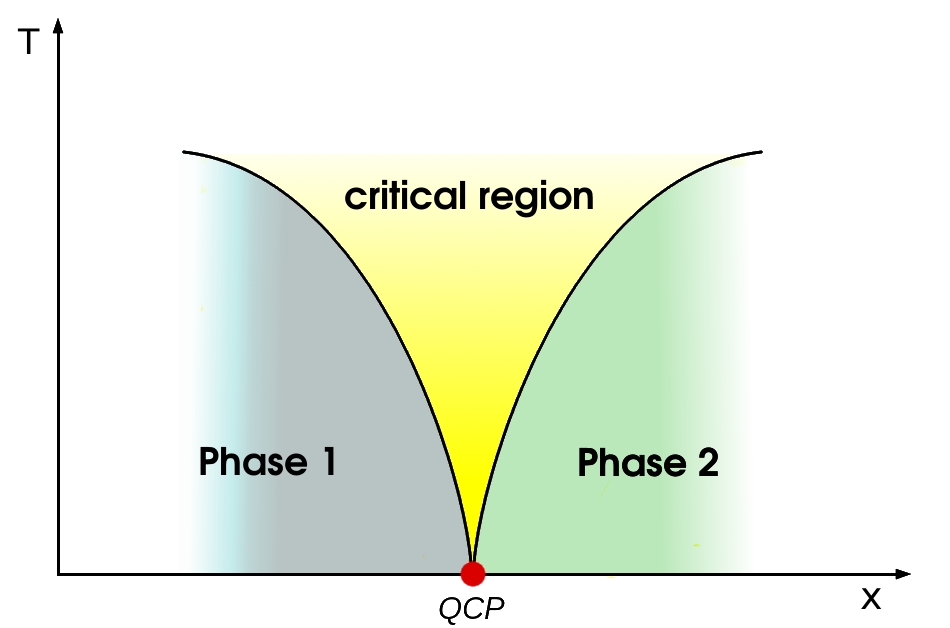} 
\caption{Cartoon representing a typical phase diagram in the vicinity of a quantum critical point.}
\label{QCP}
\end{figure}

A quantum phase transition occurs in a system at $T=0$ and at a precise point in the parameter space where
the parameters themselves attain their critical value. Such external control parameters could be, for instance, the magnetic field
or the pressure. If we move away from the quantum critical point at $T=0$ increasing the temperature we discover that the quantum critical region
widens. In other words, for low enough temperature, the extension (in parameter space) of the region of the phase diagram 
which is affected by the critical quantum dynamics increases with temperature.
At a first thought this phenomenon could sound counterintuitive, in that quantum criticality extends its relevance in the parameter space
when we move away from the quantum critical point, see Figure \ref{QCP}. 

Before attempting to have a rough intuition on why the critical region widens when we increase the temperature,
it is necessary to describe some general features of quantum criticality.
A quantum critical point separates two different phases or
two states of the system characterized by distinct orderings.
The two phases have different, long-range excitations which at the critical point require a vanishing energy 
to be excited. 
Let us indicate with $\Delta$ the energy of the lowest excited mode of the system in the vicinity of the quantum critical point. 
The value of $\Delta$ vanishes at the critical point with a power-law behavior
\begin{equation}
 \Delta \sim (x-x_c)^{\nu z}\ ,
\end{equation}
where $x$ represents some external parameter and $x_c$ is its critical value. Both $z$ and $\nu$ are positive.
We have parametrized the critical exponent of $\Delta$ in accordance with standard conventions. 
At the critical point, we have that the correlation length of the system diverges as
\begin{equation}
 \xi \sim (x-x_c)^{-\nu }\ .
\end{equation}
Comparing the two critical scalings of $\Delta$ and $\xi$, we have the standard relation defining the dynamical exponent $z$, namely
\begin{equation}
 \Delta \sim \xi^{-z}\ .
\end{equation}

Let us now try to grasp why the critical region widens when one increases the temperature
above $T=0$. The system is gapped except at the quantum critical point. At strictly null $T$
we experience criticality only where the excitations require a vanishing energy to be excited.
Moving to a non-vanishing temperature is equivalent to add a thermal noise with a characteristic 
energy $\epsilon \sim k_B T$. As a consequence, a fluctuation with finite energy
smaller than $\epsilon$ can be thermally excited. An excitation near criticality but not strictly critical
can lead to critical-like behavior as long as its characteristic correlation length is long enough
(e.g. with respect to some macroscopic characteristic length of the system). In this sense, moving away from the 
critical temperature, the critical region widens. Let us notice, however, that this kind of intuitive reasoning 
is valid as long as the quantum fluctuations dominate over the thermal ones or, equivalently, as long as 
the quantum order is not spoiled completely by thermal noise.

\subsection{Addressing the strange metal behavior in high-$T_c$ superconductors}%

This subsection is mainly inspired by \cite{Polchinski:1993ii}
where the low-energy dynamics of a system featuring spinons and 
emergent gauge fields is throughly analyzed.

\begin{figure}[ht]
\centering
\includegraphics[width=110mm]{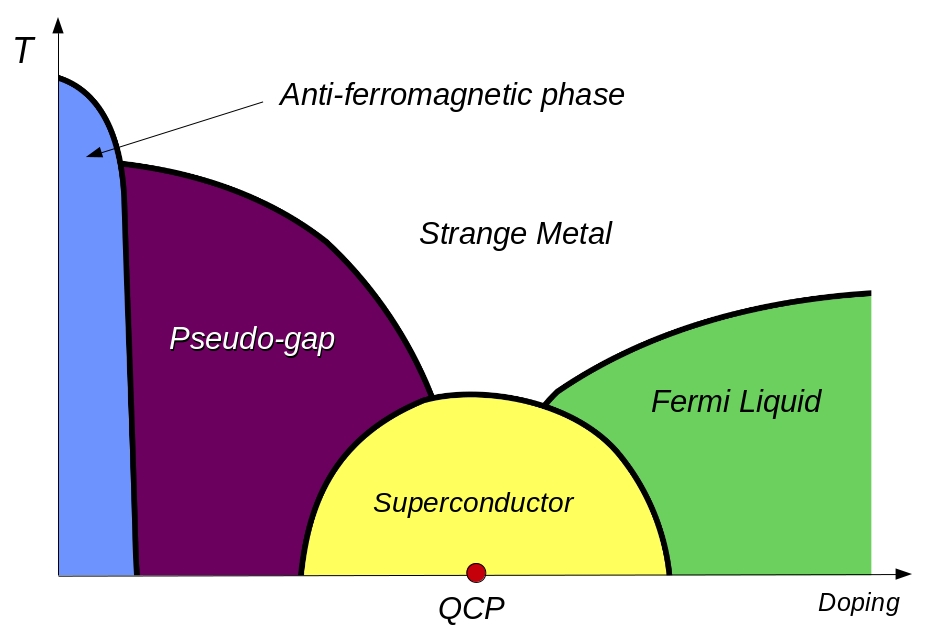} 
\caption{Cartoon of the phase diagram of a typical high-$T_c$ superconductor.
The quantum critical point is immersed in the superconducting region above which the 
system manifests strange metal behavior.}
\label{phase}
\end{figure}

The Landau-Fermi liquid is an effective field theory which describes normal metals 
but is not adequate to describe the normal phase of high-$T_c$ superconductors,
usually referred to as ``\emph{strange metal}''.
Referring to the sketchy phase diagram in Figure \ref{phase}, we focus 
on the region above the superconducting dome where actually the system manifests strange metal
behavior. Its strangeness relies on the temperature 
dependence of some of its properties which deviates significantly from the behavior
expected from the standard Landau-Fermi liquid theory. For instance, the 
resistivity of a strange metal depends linearly on $T$ instead of having a quadratic dependence%
\footnote{The strange metal behavior is manifested by many systems such as
iron pnictides and heavy fermions compounds. To have a wider description of the 
strange metal panorama see \cite{Sachdev:2011PhT} and references therein.}.
More precisely, the decay rate $\Gamma$ of a current carrier
in a strange metal (a quantity which is of course directly related to the resistivity) 
depends linearly on the biggest among $E$ or $T$ where $E$ indicates 
the excitation energy of the carrier mode itself, see \cite{Polchinski:1993ii}
for further details%
\footnote{For an explicit and holographic realization of a system 
showing linear in $T$ resistivity we refer to \cite{Donos:2012ra}. We also signal \cite{Faulkner:2010da}
for a study relating local quantum criticality and the strange metal behavior, still in 
a holographic framework. The linear in $T$ behavior has emerged also in D-brane models, 
see for instance \cite{Karch:2009eb}.}.

At low temperature, the linear in $T$ (instead of quadratic) dependence of the carrier inverse lifetime $\Gamma$ signals an enhancement of
the interactions responsible for the degradation of the current. 
Already at a naive level, one could guess that the strange metals are accordingly more
efficient than normal metals in scattering the current carriers. This, in turn, points toward 
the presence of many easy excitable degrees of freedom in the system with which
the current carrier can interact. Observe that the addition 
of new degrees of freedom interacting with the modes of the Fermi surface could 
bring the system away from the universality class of the Fermi liquid. In other 
terms, this is exactly what we look for: a system whose deep IR dynamics is not 
accountable by the Fermi liquid fixed point and its long-lived quasi-particle excitations.

To the purpose of describing the strange metals,
specific models have been proposed in which the anomalous behaviors are somehow accounted
by means of IR relevant (in the proper renormalization theory sense) modifications of the Fermi surface.
However, these modifications are typically related to some fine tuning of the effective models.
The presence of fine tuning appears to be unsuitable to unveil the mechanisms at the basis of the strange 
metal behavior. In particular, a fine tuned model seems inadequate to capture the robustness
of the strange metal behavior which, in fact, emerges on a wide parameters range (e.g. the doping,
as suggested by Figure \ref{phase}).

Before (and then independently%
\footnote{Remember that causality has to be respected!})
of $AdS$/CFT, it was proposed that the strange metal physics could be addressed
by means of a theory with a dynamically generated field. 
However, already the simplest instances of such emergent theories, turn out to 
possess a rather complicated dynamics. Especially because they are usually strongly coupled
at low-energy. It was still 1993 when Polchinski proposed \cite{Polchinski:1993ii} that the right attitude
to seek for an effective framework to describe the strange metals is to demand \emph{naturalness}.
This opposes any fine tuning and then could account for the robustness of the strange metal physics.

Let us enter in some more detail about the attempts to describe the strange metal behavior of
high-$T_c$ superconductors.
The cuprates have a Fermi surface. Experimental evidence comes from the study of quantum oscillations
in the transport properties as the applied external magnetic field is varied. Roughly speaking,
these oscillations are associated to the Landau levels traversing the Fermi surface%
\footnote{See for instance \cite{2007Natur.447..565D} and reference therein.}.
The non-Fermi character of the system is associated to the absence of quasi-particle (i.e. long-lived) excitations.
The short life of the excitations is actually in agreement with the idea that they are efficiently dissipated by interactions. 
One could then suggest to modify the standard Fermi-liquid picture by adding novel
light degrees of freedom. Goldstone bosons associated to some broken symmetry could do the job.
Quite naturally, the thought runs to phonons. Of course they are ubiquitous in condensed matter 
and strange metals are no exception; however, below the Debye temperature, they contribute to the 
relaxation of the currents only as $T^3$. As a consequence, they cannot be the cause of the non-Fermi, 
linear behavior we are interested in.

The presence of an anti-ferromagnetic region at low doping could suggest that magnetic degrees 
of freedom may enter the game as well. This naive expectation could be good in principle but it is nevertheless
based on a weak argument; note in fact that the AF phase occurs in a doping region which is far both 
from the strange metal valley and the superconducting dome (see Figure \ref{phase}). 
At any rate, the possibility of having an 
interplay between different orderings and, in particular, a superconductor and a magnetic one is very 
attractive%
\footnote{More attention will be paid on this topic in later sections}. 
Observe, for instance, that the Cooper-like pairing in high-$T_c$ superconductors is d-wave; 
in this sense, the mediators responsible for such a pairing mechanism are naturally spinful degrees of freedom.

A further possibility to reduce the carrier lifetime is to have efficient interactions with a gauge field. 
As argued in \cite{Polchinski:1993ii}, the electro-magnetic gauge field is not suitable to the purpose 
of explaining the strange metal behavior. First, the scalar potential is screened and then too short ranged; 
second, the vector potential (which is not shielded) produces effects which are suppressed by the square 
of the speed of light. The latter could in principle lead to non-Fermi liquid behavior, but in an energy range
which is much lower with respect to that which is relevant for describing strange metals. 

The presence of another gauge field besides the electro-magnetic one, either not screened or characterized 
by a lower ``speed of light'', appears as a nice possibility to address the strange metal anomalous resistivity
question. The main point is to investigate whether an effective field theory description of an emerging gauge field 
provides a suitable way to grasp the strange metal dynamics; to this end we refer to \cite{Polchinski:1993ii}. 
Here we are simply interested in showing how an emergent theory could lead us to study an effective conformal model
which, in turn, furnishes a natural context where to apply the holographic approach.

Let us begin with electrons on a lattice. In order to account for the effects of their strong Coulomb 
repulsion, we consider the constraint preventing multiple occupation of the same lattice site (by electrons of 
either spins). Note that this introduces an approximation. We describe the electrons defining the spinon 
field $f_{i\alpha}$ and the holon field $b_i$. The index $\alpha$ accounts for the spin, while $i$ labels
the lattice site. In terms of spinons and holons, the constraint forbidding double occupancy is expressed as follows
\begin{equation}\label{constra}
 \sum_{\alpha = 1}^2 f^\dagger_{i \alpha} f_{i \alpha} + b^\dagger_i b_i = 1\ .
\end{equation}
Such constraint forces to have either an electron or a hole on the site $i$; of course,
we consider an identical constraint for each lattice site.

The spinon and holon fields introduce a redundant description for the electrons. This is manifest 
as soon as we express the electron field in terms of the new fields, namely
\begin{equation}\label{ferspi}
 \psi_{i \alpha} = f_{i \alpha} b_i^\dagger\ .
\end{equation}
The electron field destroys a spinon and at the same time creates a hole%
\footnote{Notice that \eqref{constra} is mathematically equivalent to
\begin{equation}
  \sum_{\alpha = 1}^2 \psi^\dagger_{i\alpha} \psi_{i\alpha} \leq 1\ .
\end{equation}
Indeed, using \eqref{ferspi} and expressing the electron operators in terms of spinons and holons
we obtain
\begin{equation}\label{man}
 \left(\sum_{\alpha=1}^2 f_{i\alpha}^\dagger f_{i\alpha}\right) \left(1 + b^\dagger_i b_i \right) \leq 1\ ;
\end{equation}
we see that if we have an electron of either spin at the lattice site $i$ we cannot have neither
another electron nor a hole. Analogously, if we have a hole, the only way of satisfying \eqref{man} 
is to have zero electron number.}

The redundancy of the spinon/holon description consists in the possibility of considering the 
following (position dependent, i.e. local) phase transformation
\begin{equation}\label{emecha}
 f_{i\alpha}(t) \rightarrow e^{i\lambda_i(t)} f_{i\alpha}\ , \ \ \ \
 b_{i}(t) \rightarrow e^{i\lambda_i(t)} b_{i}\ ,
\end{equation}
without affecting the electron field. In light of the gauge transformation \eqref{emecha}, the constraint \eqref{constra}
is a local charge conservation requirement with respect to the charge associated to 
the emergent gauge field.

At this point it is fair to ask ourselves why we should adopt a redundant (though in some sense 
equivalent) description in terms of emergent degrees of freedom. The reason is phenomenological.
There are specific physical situations, mainly quantum critical points, where the description
in terms of emergent or ``fractionalized'' degrees of freedom is more accurate 
in the sense that it is able to capture important physical behaviors especially related to 
topological properties. More precisely, the description in terms of the new emergent degrees 
of freedom could actually differ from the original one about global or, actually, topological 
properties.

The spinon/holon system with emergent gauge symmetry is a prototypical
example of a theory with massless degrees of freedom. Indeed, the gauge boson 
masslessness is related to the gauge invariance. Notice that technically the 
gauge structure emerges because we want to implement a constraint on a redundant description.
On a phenomenological level, the gauge boson accounts for critical collective excitations of the system.
Indeed the emergent gauge theory picture applies at a (fractional) critical point. 
Intuitively, it can be thought as a generalization of what is standard in Ginzburg-Landau theory 
where, at criticality, the fluctuations of an order parameter become massless and their 
correlation length diverges.

%
%
%
%

\subsubsection{Comment about the relation between Fermi and non-Fermi liquids}

The Landau-Fermi liquid picture proves to be extremely powerful in accounting
for many metallic materials and furnishes the basis of the study of many
interesting properties. For instance the BCS theory for superconductivity.
Of course, it is not surprising that, as soon as the Landau-Fermi liquid picture
is not valid and, in particular, when the lifetime of the quasi-particle excitations
becomes small, we are immediately faced with a tremendously complicated problem.
Namely, a many-body system where we do not have long-lived modes and where, consequently,
it is troublesome to build any perturbative theoretical construction. 

Somehow the Landau-Fermi picture is standard and we are acquainted to think in 
terms of its framework. However, the fact that it works is a priori pretty astonishing.
When attempting to describe a metal, we face a microscopic (said otherwise, UV) 
many-body system where the electrons interact strongly with the lattice. 
However, thanks to the Pauli exclusion principle and the formation of a Fermi surface,
the relevant modes are those corresponding to the momentum shell close to the 
Fermi surface itself. Such modes need a small amount of energy to be excited; note
indeed that the structure of the Fermi surface itself produces a suppression of the phase 
space of the modes.
In other terms, the presence of the Fermi sea reduces the efficiency of the 
quasi-particle scattering.
Such constraint in the phase space is intuitively the reason why the interactions 
of the quasi-particle are usually attenuated and admit a weak coupling description.
More precisely, in a Renormalization Group terminology, we say that the Fermi liquid
is the IR fixed point of the complicated electron/lattice system. Such IR fixed 
point features long wavelength excitations whose weakly coupled dynamics is insensitive 
to the detailed UV (strong) interactions. Said otherwise, such interactions are IR irrelevant.

As a final comment, it is fairer to be surprised by the simplicity of the Landau-Fermi
picture rather than the complexity of the non-Fermi systems!

\subsection{Minimal ingredients to superconduct}
\label{DCinf}

It is important to pinpoint which are the essential symmetry features of a gauge theory which 
lead to the phenomenological properties of a superconductor, in particular, a diverging
DC conductivity. Actually, we start by assuming that the superconducting medium allows 
a description in terms of a quantum gauge field theory. Specifically, a $U(1)$ theory 
where the gauge group coincides with the usual electro-magnetic Abelian invariance.
We therefore consider an action that is invariant under the following gauge transformations:
\begin{eqnarray}\label{gauge}
 & A_\mu(x) & \rightarrow A_\mu(x) + \partial_\mu \alpha(x) \ , \\
 & \xi(x)  & \rightarrow e^{i q \alpha(x)} \xi(x) \ .
\label{rigauge}
\end{eqnarray}
Note that in writing explicitly the gauge transformation we have tacitly assumed the presence 
of a single species of charged fermions $\xi$ with electric charge $q$.
$\alpha(x)$ is the arbitrary gauge parameter function which specifies the particular gauge 
transformation we want to consider.
If we pick an individual point $\overline{x}$ in space-time, the gauge transformations 
\eqref{gauge} and \eqref{rigauge} correspond to a compact $U(1)$ phase symmetry;
actually, the values $\alpha(\overline{x})$ and $\alpha(\overline{x}) + 2\pi/q$ are 
accordingly identified.

Generically, in describing a superconductor, the gauge symmetry \eqref{gauge} is supposed 
to be broken by the spontaneous condensation of some charged operator.
Let us suppose that, in the broken phase, the original local $U(1)$ symmetry is reduced 
so that only a discrete subgroup $\mathbb{Z}_n\in U(1)$ remains preserved. 
Goldstone's theorem states that the spontaneous symmetry breaking causes the appearance 
of a massless mode $G$ which parameterizes the coset group $U(1)/\mathbb{Z}_n$. 
The Goldstone field $G$ behaves as a phase and then its gauge transformation is
\begin{equation}\label{Gol_gau}
 G(x) \rightarrow G(x) + \alpha(x) \ .
\end{equation}
Furthermore, as the Goldstone boson spans the coset group $U(1)/\mathbb{Z}_n$, we have the following identifications
\begin{equation}
 G(x) = G(x) + \frac{2\pi}{n\, q}\ .
\end{equation}

Relying on gauge invariance arguments we have that the Lagrangian for the theory describing 
the gauge and Goldstone's fields needs to have the following general structure
\begin{equation}\label{laggau}
 {\cal L} = -\frac{1}{4} \int d^dx \ \left\{ F \cdot F + L_G [A - d G] \right\} \ ,
\end{equation}
where the detailed form of the Goldstone part $L_G$ of the Lagrangian density depends on the specific model.
On the contrary, the functional dependence of $L_G$ on $A - dG$ is a general feature descending from gauge 
invariance requirements. Note indeed that, according to the transformations \eqref{gauge}, $A-dG$ is a gauge 
invariant quantity.

Consider the system described by the Lagrangian \eqref{laggau}.
The electric current and the charge density are given by\footnote{Note that in the present
treatment we are assuming Euclidean space-time.}
\begin{eqnarray}
 & J^i & = \frac{\delta L_G}{\delta A_i(x)} \ , \\
 & J^0 & = \rho = \frac{\delta L_G}{\delta A^0(x)} = - \frac{\delta L_G}{\delta (\partial_t G)} \ . \label{chaden}
\end{eqnarray}
In the last passage we have used the assumption that the matter part of the Lagrangian density
(i.e. $L_G$) depends only on the gauge invariant combination $A-dG$.
Observe that Equation \eqref{chaden} implies that $-\rho$ is the canonical conjugate variable to $G$.
This, in turn, means that, within a Hamiltonian description, the energy density $\cal H$ is 
expressed as a functional depending on $G$ and $\rho$.
We can therefore write the Hamilton equation for $\partial_t G$ which is
\begin{equation}\label{hami}
 \partial_t G(x) = - \frac{\delta {\cal H}}{\delta \rho(x)}\ .
\end{equation}

The physical interpretation of the Hamilton equation \eqref{hami} allows us to 
discover that the system at hand possesses the defining property of a superconductor, 
namely a diverging D.C. conductivity.
The Hamiltonian ${\cal H}$ represents the energy density of the system while $\rho(x)$ denotes the charge density.
As a consequence, the right hand side of \eqref{hami} expresses the variation in energy density due to a variation
of the charge density. In other terms, the electric potential.
We reach the conclusion that the time derivative of the Goldstone field $G$ is then related to the potential
in the following manner:
\begin{equation}\label{pote}
 \partial_t G(x) = -V(x) \ .
\end{equation}
Take now a stationary state and assume that there is a steady current flowing through the superconducting medium.
The stationary assumption implies that nothing depends explicitly on time and hence we have $\partial_t G(x) = 0$.
Relying on Equation \eqref{pote}, we see that the electric potential $V(x)$ is then forced to be vanishing as well.
As we have a stationary electric current without having any difference of potential sustaining it, 
we are dealing with a system characterized by a zero resistance or, equivalently, infinite conductivity.
Moreover, since we are here focusing on the stationary properties of the system, we are actually studying its
DC conductivity. The DC conductivity is defined in terms of the optical conductivity by the following limit
\begin{equation}
 \sigma_{\text{DC}} = \lim_{\omega\rightarrow 0}\sigma(\omega) \ .
\end{equation}

We have just described the possibility of having infinite DC conductivity basing our 
arguments (originally suggested by Weinberg in \cite{Weinberg:1986cq}) on very simple assumptions:
\begin{itemize}
 \item The presence of an Abelian electro-magnetic gauge symmetry
 \item Its spontaneous breaking down to a discrete subgroup
\end{itemize}
No precise detail of the actual breaking mechanism have been specified. 
The message to take home is the recognition of the importance of the spontaneous symmetry breaking 
itself in producing the fundamental phenomenological features defining superconductivity.
Even without any precise reference to the microscopic origin of the breaking mechanism itself.
In this perspective, we can better understand the reason why phenomenological models \emph{\'a la} Ginzburg-Landau
are actually capable of describing accurately the phenomenological features of superconductivity even though they 
rely on rough approximations such as the description of the Cooper pairs with a single bosonic field.

Later on in these lectures we will illustrate how a similar effective and minimal approach constitutes the
basis of the holographic description of superconducting systems.

\section{Holographic superconductor: the minimal model}

Here we introduce the gravity model which is dual to a boundary, strongly coupled theory
describing a superconductor. Being the gauge/gravity correspondence a strong/weak duality,
we are interested in the weak interacting regime of the gravity model. As it is standard in 
holography, one considers the gravity model in the limit where the typical size of the geometry
is far bigger than the typical string length and where one can neglect quantum corrections.
The bulk theory is accordingly studied in a semiclassical approximation and, on a practical level,
we are concerned in solving and studying the classical equations of motion of the gravity model. 
Exploiting then the standard holographic 
dictionary, the semi-classical results obtained from the bulk model are read and interpreted 
in terms of correlation functions of the strongly coupled, quantum theory living at the boundary%
\footnote{For further details we refer to the classic $AdS$/CFT literature, in particular to 
\cite{Aharony:1999ti,Zaffaroni:2000vh}.}.

\hspace{10pt}

\noindent
{\bf The field content:} The bulk model contains gravity. The bulk graviton is dual to the boundary 
energy-momentum tensor and then it is ubiquitous in any holographic model. In relation 
to the gravity part, we consider standard Einstein-Hilbert action with a negative cosmological
constant so that to have $AdS$ vacuum solutions. Let us remind that the bulk geometry
accounts for crucial properties of the dual boundary system. In particular the presence of a
horizon (i.e. a black hole) in the bulk, associated to a finite value of the Hawking temperature, 
corresponds to a finite temperature boundary quantum field theory%
\footnote{More details on the correspondence of the bulk and boundary thermodynamics
are given in the following sections.}.

The superconductor phenomenology necessitates the presence of charged degrees of freedom 
whose condensation leads to superconductivity. In the holographic framework, the charged degrees of freedom
are accounted for by means of a chemical potential. Such chemical potential of the boundary theory 
is dual to the temporal component of a bulk gauge field. Notice that a bulk geometry presenting a non-trivial
profile for the gauge field corresponds to a charged solution, typically a charged black hole.

Eventually, we need some degrees of freedom to describe the charge condensate. The superconductors (also the 
conventional ones) are classified according to the rotation symmetry of the superconducting order parameter.
The s-wave superconductor has an isotropic (spin $0$) condensate which is naturally modeled by a charged scalar operator.
In holography, the tensor properties of the boundary operator is the same as that of the dual bulk field.
Indeed one must be able to saturate the boundary operator with the boundary value of the bulk field to build
an invariant (i.e. scalar) source term for the boundary action. The bulk field allowing us to study a scalar condensate 
is therefore a charged (with respect to the bulk gauge field) scalar field. Finite temperature bulk solutions possessing a non-trivial 
profile for such a scalar field are hairy black holes and corresponds dually to configurations of the superconductor where
the superconducting condensate is not vanishing.

\hspace{10pt}

\noindent
{\bf The bulk action:} Having specified the field content, we introduce an action for the model.
Since we are following a bottom-up approach, we choose the simplest action to begin with.
We indeed consider minimal coupling in two senses, both gravitationally and electro-magnetically.
The matter fields, i.e. the gauge vector and the charged scalar field, are minimally coupled to 
gravity and this latter is minimally coupled with the electro-magnetic field.

We then consider the Abelian Higgs model in 4 dimensions with Einstein-Hilbert gravity
described by the action
\begin{equation}\label{actia}
 S = \frac{1}{2 \kappa_4^2} \int dx^4 \sqrt{-g}
 \left\{ {\cal R} + \frac{6}{L^2} - \frac{1}{4}F_{ab}F^{ab}
 - V(|\psi|) - |\partial \psi - i q A \psi|^2 \right\}\ ,
\end{equation}
where $\kappa_4$ is the 4-dimensional gravitational coupling.
The term $6/L^2$ represents a negative cosmological constant. 
The model admits $AdS_4$ solutions without matter and
asymptotically $AdS_4$ solutions when matter is present%
\footnote{The near-boundary $AdS_4$ is dual to a conformal UV fixed point.
This being the reason for us to choose the asymptotically $AdS$ solutions.
Nevertheless, other solutions with different near-boundary behaviors do exist.}
; the $AdS_4$ radius of curvature is $L$.
Insofar we have kept $L$ explicit because it is useful for the dimensional analysis;
however, we will later consider $L=1$. Fixing $L$
to be unitary corresponds to a choice of units of length and, as we will see studying
the scaling invariance of the equations of motion descending from the action \eqref{actia},
such a choice of units does not spoil the generality of the description.

The choice of a $4$-dimensional bulk is phenomenologically inspired. Many real-word high-$T_c$
superconductors (such as the cuprate systems) possess a layered structure and unconventional 
superconductivity is believed to occur due to intra-layer dynamics. To a first approximation,
in order to study the dynamics producing superconductivity, it is enough to focus on a single layer.
This has $2$ spatial dimensions as the boundary of a $3+1$ bulk.

\hspace{10pt}

\noindent
{\bf Bulk dimensional analysis:} With bulk dimensional analysis we mean that all $4$
bulk coordinates are considered on the same footing. Specifically, the radial coordinate
$r$ is here a bulk spatial coordinate with the same dimensions as $x,y$ and $t$.
The gauge field $A$ and the scalar field $\psi$ are both 
dimensionless. The charge $q$ has the same dimension of $\partial$, then, an inverse length.

\hspace{10pt}

\noindent
{\bf Coupling constants:} We have the coupling constant $q$ which represents 
the charge of the complex scalar $\psi$. Then, there is the $4$-dimensional
gravitational coupling $\kappa_4$ which nonetheless can be fixed to $1$
without loss of generality. Indeed, it represents an overall normalization 
of the action. The potential $V$ may introduce new couplings and new parameters 
to the system. Keeping a generic potential $V$ is in the spirit of a bottom-up approach%
\footnote{A ``radical'' example of bottom-up model where phenomenological function coefficients 
are introduced also in the kinetic term of the vector field is given in \cite{Aprile:2009ai,Aprile:2010yb}. 
This procedure of introducing functions to be phenomenologically determined is referred to as ``holographic fit''.}. 
In fact, in a UV consistent framework the potential can be constrained by the specific UV
completion which one considers. On a practical level, we will focus the attention mainly on the simplest 
non-trivial choice for the potential, namely
 \begin{equation}\label{potma}
  V(|\psi|) = m^2 |\psi|^2 + ...\ ,
 \end{equation}
 where $m$ represents a mass for the scalar and its dimension is $[m]= [L]^{-1}$.
 The value of the mass is an important feature. Not only because it represents a parameter 
 upon which the equilibrium properties of the system depend (see for instance \cite{Horowitz:2008bn}), but also in relation to the 
 stability of the background. More comments on this latter aspect are given in Subsection \ref{scastab}.

\subsection{Phenomenological interpretation}

At first, one could wonder how we mean to account for superconductivity
without having any fermion around in the model. Actually, the description 
we give is in the effective spirit, where the quantity we focus on is rather 
a bosonic condensate of some multi-fermion bound state. 
In this case, the effectiveness of the approach relies in dealing with the condensate degrees 
of freedom instead of handling its microscopic constituents.

Of course, in the holographic model \eqref{actia}, the condensate is encoded
with the complex scalar field $\psi$; its charge being related to the charge 
of the composite fermion state%
\footnote{Such composite fermionic state is in principle a generalization of a Cooper-like pair.
As we will comment in the following, the holographic superconductors hint
towards the possibility of generalizing (in a strongly coupled context) 
the Cooper pairing mechanism; yet, the number of microscopic participants to the 
bosonic bound state can be different than $2$. See Subsection \ref{carrier}.}.
The scalar nature of the condensate field $\psi$ corresponds to the fact that 
the bound electrons give rise to an overall spinless state. This is the 
reason why the holographic model at hand is referred to as an s-wave 
superconductor. Thinking of a Cooper-like pairwise mechanism, the electrons 
would be coupled in a singlet spin state. Observe that these rough sentences
are strictly speaking illegitimate at strong coupling, where no talking of single 
or coupled electrons is technically permitted. However, let us adopt this language 
to guide the intuition and with the caveat that, although we are inspired 
by the weak-coupling language, we are dealing with a system which is intrinsically 
different and describable in terms of some other set of degrees of freedom.

\subsection{Equations of motion}

From the action \eqref{actia} we have the scalar 
equation of motion
\begin{equation}\label{equsca}
 -(\nabla_a - iqA_a) (\nabla^a-iqA^a) \psi 
 + \frac{1}{2} \frac{\psi}{|\psi|} V'(|\psi|) = 0\ ,
\end{equation}
the Maxwell equations
\begin{equation}\label{maxsca}
 \nabla^a F_{ab} = iq [\psi^*(\nabla_b-iqA_b)\psi - \psi(\nabla_b+iqA_b)\psi^*]\ ,
\end{equation}
and eventually the Einstein equations
\begin{equation}\label{Ei}
 \begin{split}
 &R_{ab} - \frac{g_{ab} R}{2} - \frac{3 g_{ab}}{L^2} = 
 \frac{1}{2} F_{ac}F_b^{\ c} - \frac{g_{ab}}{8} F^{cd}F_{cd}
 - \frac{g_{ab}}{2} V(|\psi|) - \frac{g_{ab}}{2}|\nabla \psi - i q A\psi|^2\\
 &+\frac{1}{2}[(\nabla_a\psi-iqA_a\psi)(\nabla_b\psi^*+iqA_b\psi^*)
 +(\nabla_b\psi-iqA_b\psi)(\nabla_a\psi^*+iqA_a\psi^*)]\ .
 \end{split}
\end{equation}
The equations of motion give rise to a pretty complicated system. We study that with 
a simplifying ansatz where all the fields present just radial dependence. Notice that,
adopting such an ansatz, the system reduces to a set of ordinary differential equations.
In order to study spatial features or the time evolution one has to generalize the ansatz.
This could bring a rather crude increase in the technical difficulty, especially on the 
computational level. Nevertheless, already a simple radial ansatz allows us to unveil
many features of the behavior of the holographic superconductor. We will spend some attention 
on spatial and temporal dependent solutions when describing the generalization of the
minimal holographic superconductor.

\vspace{10pt}

\noindent
{\bf Ansatz:} In formul\ae,\ the radial ansatz we consider is given by
\begin{equation}\label{gans}
 ds^2 = - g(r) e^{-\chi(r)}\, dt^2 + \frac{dr^2}{g(r)} + r^2 (dx^2 + dy^2)
\end{equation}
and
\begin{equation}\label{phipsians}
 A = \phi(r) dt\ , \ \ \ \
 \psi = \psi(r)\ .
\end{equation}
It is easy to see that our ansatz contains the $AdS_4$
solution. Requiring asymptotic $AdS$-ness of the solutions
is equivalent to demand the following UV behavior for the metric 
coefficient functions,
\begin{equation}\label{asyAdS}
  g(r)\ \overset{r\rightarrow\infty}{\longrightarrow}\ r^2 + ...\ , \ \ \ \
  \chi(r)\ \overset{r\rightarrow\infty}{\longrightarrow}\ 0 + ...
 \end{equation}
 
\vspace{10pt}
\noindent 
{\bf Gauge choice:} Consider the bulk perspective. We have a model with a 
dynamical gauge field which is treated classically. In order to 
perform explicit computations we have to choose a gauge. Of course, all 
bulk physical quantities are insensitive to such a choice.
We consider the axial gauge
\begin{equation}\label{axial}
 A_r = 0\ .
\end{equation}
The ansatz \eqref{phipsians} is actually far more restrictive 
but, for sure, it implies the gauge condition \eqref{axial}.
Having chosen the axial gauge, the $r$ component of Maxwell's equations 
fixes the phase of the scalar field $\psi$ to be constant. 
We are then allowed to take $\psi$ to be real. 
Note that the gauge transformations on $\psi$ rotate it without 
changing its modulus. The results that we obtain in the gauge
where the field $\psi$ is real, are therefore valid in general in relation 
to the modulus $|\psi|$. In other terms, even working in a specific gauge,
the modulus of the field $\psi$ is a gauge invariant quantity and it is 
in terms of it that we describe the superconductor physics we are interested in.

As we will see later in Subsection \ref{stu},
given our gauge choices, the bulk electro-magnetic current is expressed as
$J_\mu \sim |\psi|^2 A_\mu$. Being both the electro-magnetic current 
and the modulus of $\psi$ gauge invariant, we conclude that the 
explicit values of $A_\mu$ we will encounter in our gauge are simply 
related to gauge invariant quantities.

\vspace{20pt}

\noindent
{\bf Explicit system of equations of motion after imposing the radial ansatz}

\vspace{10pt}
\noindent
The equation of motion \eqref{equsca} for the scalar field $\psi$ becomes
\begin{equation}\label{eqscal}
 \psi'' + \left(\frac{g'}{g} - \frac{\chi'}{2} + \frac{2}{r} \right) \psi'
 + \frac{q^2 \phi^2 e^\chi}{g^2} \psi
 - \frac{1}{2g} V'(\psi) = 0 \ .
\end{equation}
Apart from the ``bare'' mass $m$ contributed by the quadratic 
term in $V$ (see Equation \eqref{potma}), the scalar field acquires an effective 
mass given by
\begin{equation}
 \tilde{m}^2_\psi = m^2 - \frac{q^2 \phi^2 e^\chi}{g}\ .
\end{equation}
Note that the sign of the electro-magnetic contribution to the scalar mass
is crucial in order to lead the system towards an instability. Both the charge $q$ and the 
field $\phi$ are squared. The function $g$ is always positive, indeed $g$ is positive at the 
boundary and it cannot have but one zero. We remind the reader that $g(r)=0$ 
actually signals the presence of the horizon (see Subsection \ref{stu}).

Considering the radial ansatz, the Maxwell equation \eqref{maxsca} becomes
\begin{equation}\label{maxequ}
 \phi'' + \left(\frac{\chi'}{2} + \frac{2}{r} \right) \phi'
 - \frac{2 q^2 \psi^2}{g} \phi = 0\ .
\end{equation}
From the structure of the equation we understand that, whenever the 
scalar field $\psi$ is non-trivial, the temporal component $\phi$ 
of the gauge field acquires a mass according to a classical version of
the bulk Brout-Englert-Higgs mechanism. Specifically, from the equation
we read
\begin{equation}
 \tilde{m}^2_\phi = 2 q^2 \psi^2\ .
\end{equation}

We have two further equations for the metric functions $g$ and $\chi$ which assume the following form
\begin{equation}\label{eq_chi}
 \chi' + r \psi'^2 + \frac{r q^2 \phi^2 \psi^2 e^\chi}{g^2} = 0\ ,
\end{equation}
\begin{equation}\label{equaps}
 \frac{1}{2} \psi'^2 + \frac{\phi'^2 e^\chi}{4g}
 + \frac{g'}{g r} + \frac{1}{r^2}
 - \frac{3}{g L^2} + \frac{V(\psi)}{2 g}
 + \frac{q^2 \psi^2 \phi^2 e^\chi}{2 g^2} = 0\ .
\end{equation}
Note that these last two equations are first-order as opposed to the previous equations for the scalar
and gauge fields which are second-order.

\vspace{10pt}

\noindent
{\bf Probe approximation:} In the probe approximation we consider 
the matter fields (i.e. the scalar and the gauge fields)
as small perturbations on the gravitational background
which remains insensitive to them. In other terms, we neglect the backreaction 
of the gauge and scalar fields on the bulk geometry. Of course such an approximation
is meaningful if the profiles of the matter fields are in some sense (to be specified precisely) 
small enough. From the physical bulk perspective, the probe approximation
is accurate as long as the energy density of the matter fields is negligible
with respect to the energy density of the geometry itself.
Note that this ``energetic'' viewpoint is meaningful also from the boundary theory
perspective. Consider the fluctuations of the bulk metric; they account in a dual 
fashion for the energy and momentum transport of the boundary model%
\footnote{The study of the linear response and transport of the holographic
superconductor constitutes the subject of Section \ref{trans}.}. In a probe
framework, the dynamics of the metric is insensitive to the matter fields. Hence these latter 
do not contribute to the transport of energy and momentum. Said otherwise, the probe approximation
works whenever their energy/momentum contribution is negligible.

A direct study of the equations of motion allows us to define neatly
the probe approximation. It is indeed easy to observe that, if we consider the 
large charge limit
 \begin{equation}
  q \rightarrow \infty \ ,
 \end{equation} 
while keeping $q \psi$ and $q\phi$ fixed, the terms containing
the matter fields $\psi$ and $\phi$ become negligible in the equations
for the metric functions $\chi$ and $g$ (see \eqref{eq_chi} and \eqref{equaps}).
To rephrase, this means that $\chi$ and $g$, and then (according to \eqref{gans}) the metric as a whole, are
insensitive to the matter fields.
 
After having discarded the $\phi$ and $\psi$ dependent terms in the equations 
for $\chi$ and $g$, they admit as a solution the Schwarzschild black hole, namely
 \begin{equation}
  \chi(r) = 0\ , \ \ \ \
  g(r) = r^2 -\frac{r_h^3}{r} \ .
 \end{equation}

\subsection{Studying and solving the system of e.o.m.}
\label{stu}

\vspace{10pt}
\noindent
 {\bf Black hole horizon:} The black hole horizon is a stationary and null surface. These features
 imply that the horizon causally separates the exterior and interior regions \cite{Dabholkar:2012zz}.
 The stationarity can be more precisely stated requiring that the time-like killing vector 
 leaves the horizon invariant. 
 
 Given the metric ansatz \eqref{gans}, a surface defined by a fixed value $r_*$ for the radial
 coordinate is automatically null if $g^{rr}(r_*)=g(r_*)=0$; furthermore, according to the ansatz 
 \eqref{gans}, the condition $g(r_*)=0$ implies also $g_{tt}(r_*)=0$ and, in the present context, 
 it is the definition of a black hole horizon. 
 In addition, the horizon radius can be found solving the equation
 \begin{equation}\label{horco}
  g(r_h) = 0\ .
 \end{equation}

\vspace{10pt}  
\noindent
{\bf Temperature:} Focus on a generic asymptotically $AdS$ black hole whose $t,r$ parts of the
metric have the following generic shape
\begin{equation}\label{gen_sha}
 ds^2 = - a(r) \, b(r)\ dt^2 + \frac{dr^2}{b(r)} + ...
\end{equation}
where the dots stand for the pieces involving other coordinates.
We pass to Euclidean space-time obtaining
\begin{equation}\label{eumem}
  ds^2_{\text{Eucl}} =  a(r)b(r)\ d\tau^2 + \frac{dr^2}{b(r)} + ...
\end{equation}
We intend to compare \eqref{eumem} in the vicinity of the horizon with polar coordinates on the plane, namely
\begin{equation}\label{piano}
 ds^2_{\text{pol}} = dr^2 + r^2\ d\vartheta^2\ .
\end{equation}
To this end, we expand the function $b(r)$ recalling that, in agreement with \eqref{horco}, 
we have $b(r_h)=0$,
\begin{equation}\label{bdr}
 b(r) = b'(r_h) (r-r_h) + ...
\end{equation}
We consider the following change of variable
\begin{equation}\label{chax}
 \zeta^2 = \frac{4}{b'(r_h)} (r-r_h)\ ,
\end{equation}
so that
\begin{equation}
 \frac{dr^2}{b(r)} \sim d\zeta^2\ ,
\end{equation}
in the near horizon. Moreover, plugging \eqref{chax} into \eqref{bdr} 
we have, still in the near horizon region, that
\begin{equation}
 b(r) = \left( \frac{b'(r_h)}{2} \right)^2 x^2 + ...
\end{equation}
Therefore the metric takes the IR asymptotic form
\begin{equation}\label{methor}
 ds^2 = \left( \frac{b'(r_h) a^{1/2}(r_h)}{2} \right)^2 \zeta^2\, d\tau^2 + dx^2 + ...
\end{equation}
It is customary to define $\kappa = b'(r_h) a^{1/2}(r_h) /2$ which corresponds to the surface gravity.
Comparing \eqref{methor} with the polar coordinates on the plane \eqref{piano} and asking for no conical singularity
at the origin, we require the Euclidean time $\tau$ to have the periodicity
\begin{equation}
 \vartheta = \vartheta + 2\pi\ , \ \ \ \
 \tau \sim \tau + \frac{4\pi}{ b'(r_h) a^{1/2}(r_h)}\ .
\end{equation}
Recall that the inverse period of the Euclidean time gives the temperature.
We have then
\begin{equation}\label{tempp}
 T = \frac{1}{4\pi}  b'(r_h) a^{1/2}(r_h) \ .
\end{equation}
The requirement of non-singular behavior of the metric at the horizon has 
therefore provided us an explicit expression for the temperature of the 
bulk system and the bulk temperature is identified with the temperature 
of the boundary theory. One direct way to convince ourself of the identification 
of the bulk and boundary temperature comes from the observation that the time
coordinate is shared by the bulk and boundary manifolds and that the temperature 
is introduced by means of an analytic continuation and compactification of the 
Euclidean time coordinate.
 
Let us specify the general formula \eqref{tempp} to the case of the holographic 
superconductor whose action is \eqref{actia}. We express $g'(r_h)$ in terms of the other 
fields using the near-horizon expansion of \eqref{equaps} and obtain
\begin{equation}\label{temp}
 T = \frac{1}{4\pi}  g'(r_h) e^{-\chi(r_h)/2}
   = \frac{r_h}{16\pi L^2} \left[ (12 + 2 m^2 L^2 \psi^2) e^{-\chi/2} - L^2 \phi'^2 e^{\chi/2}\right]_{r=r_h}\ .
\end{equation}

\vspace{20pt}
\noindent
{\bf Physical consistency requirements for the bulk model:} We need to require that the scalar potential (i.e. the temporal component of the 
  electro-magnetic gauge field) vanishes at the black hole horizon, namely
  \begin{equation}
   A_t(r_h) = \phi(r_h) = 0\ .
  \end{equation}  
  Recall that the $tt$ component of the metric vanishes at the horizon as well. This 
  latter fact implies that the norm of a vector potential with $A_t(r_h) \neq 0$ (and either finite or null values 
  for its spatial components) has a diverging norm at the horizon, indeed
  \begin{equation}\label{infnor}
   g^{tt}(r_h) A_t(r_h) A_t(r_h) \rightarrow \infty\ .
  \end{equation}
  The quantity $A_t(r_h)$ is a priori not gauge invariant and hence possibly not physical. In principle 
  there could be no clear physical problem in having a diverging norm for a non-physical quantity.
  Nonetheless, let us observe that \eqref{infnor} leads to diverging physical quantities too. We make two examples.
  
  Consider the charge current associated to the scalar field $\psi$, namely
   \begin{equation}\label{cursca}
   \begin{split}
   J_\mu &\propto \psi^* D_\mu \psi - \psi (D_\mu \psi)^*\\
         &= \psi^* \partial_\mu \psi - i q\; \psi^* A_\mu \psi - \psi \partial_\mu \psi^* - i q\; \psi^* A_\mu \psi\ .
   \end{split}
  \end{equation}  
  We have chosen a gauge where the scalar field is real. The expression of the bulk
  electro-magnetic current \eqref{cursca} reduces to
 \begin{equation}
  J_\mu \propto i q \; A_\mu \psi^2\ .
 \end{equation}
  In our gauge, having $A_t(r_h) \neq 0$ would then lead to the divergence of the current $J_\mu$ which
  is a physical quantity. Although the argument about the need of posing $A_t(r_h)=0$ has been made here
  in a specific gauge, since we dealt with a gauge invariant current, the valence of the conclusion is 
  general and gauge independent.
  
  We can consider another gauge independent quantity, namely the Wilson loop around the compact
  Euclidean time circle. If $A_t(r_h) \neq 0$ we have a non-vanishing time-like Wilson loop at the horizon.
  However, this time-like Wilson loop at the horizon has a vanishing measure. Having a non-zero Wilson loop
  for a contour with vanishing measure implies a singularity of the Maxwell field.

  \vspace{20pt}
  \noindent
  {\bf UV fall-off's:} from the study of the near-boundary limit of the gauge and scalar 
  equations of motion \eqref{maxequ} and \eqref{eqscal}, we can derive the large $r$ asymptotic
  behaviors of the fields. We obtain
  \begin{equation}\label{asyphi}
   \phi = \mu - \frac{\rho}{r}+... \ ,
  \end{equation}
  for the gauge field and
  \begin{equation}\label{asypsi}
   \psi = \frac{\psi_1}{r} + \frac{\psi_2}{r^2} + ...\ ,
  \end{equation}
  for the scalar. Notice that these results depend on the dimensionality of the problem. 
  To have a comparison between the $2+1$ boundary and the $3+1$ boundary case we refer to 
  \cite{Horowitz:2008bn}.
  
   \vspace{20pt}
  \noindent
  {\bf Holographic dictionary%
  \footnote{In this Paragraph we follow the lines of \cite{Marolf:2006nd} to which we refer for a more detailed treatment.}:}
  The UV asymptotic behavior of the bulk fields (and specifically the power laws with respect to the radial 
  coordinate) determines the conformal dimension of the corresponding dual quantities. 
  The semi-classical treatment of the bulk model requires a well-defined variational problem. 
  Namely, we need a finite symplectic structure in the bulk and a vanishing symplectic flux at the boundaries \cite{wald1994quantum}
  (i.e. both at the conformal boundary corresponding to radial infinity and at the horizon).
  Such requirements define which are the possible boundary condition we are allowed to consider for the fields.
  In turns, the boundary conditions fixe the UV fall-off's of the bulk fields and, in a gauge/gravity framework,
  define the operator content of the CFT at the conformal boundary.
  
  In relation to the bulk gauge field $A_\mu$ whose near-boundary expansion is given in \eqref{asyphi}, 
  we consider the Dirichlet boundary conditions fixing the leading term $\mu$ to a constant%
  \footnote{To have a review of other possibilities and their relevance in the context of gauge/gravity duality see \cite{Marolf:2006nd}.}.
  Doing so, we define a framework where the leading term $\mu$ in the near-boundary expansion is interpreted as 
  a source of the CFT while the subleading term $\rho$ instead corresponds to the VEV of the associated operator
  (i.e. that sourced by the $\mu$ term).
    
  To fully appreciate the physical interpretation of the UV terms as a chemical potential 
  and a charge density (as suggested by the notation introduced in \eqref{asyphi})
  we have to recall how the gauge/gravity correspondence enforces the identification of the
  field theory sources with the gravity theory boundary values of the bulk fields. On the field theory 
  side, a vector source $A_\mu$ is coupled to a vector current $J_\mu$ through a term of the following type
  \begin{equation}\label{sou_ter}
   \int d^{d}x\ A_{(0)}^\mu J_\mu\ , 
  \end{equation}
 where $A^\mu_{(0)}$ is the boundary value of the bulk gauge field. According to the ansatz 
 \eqref{phipsians}, we have that only the temporal component of the gauge field is non-null.
 Recall that the temporal component of a current densisty is a charge density, so \eqref{sou_ter}
 becomes
  \begin{equation}\label{sou_ter_tem}
   \int d^{d}x\ A_{(0)}^0 J_0 = \int d^{d}x\ A_{(0)}^0 \rho\ .  
  \end{equation}
  Still in agreement with our ansatz \eqref{phipsians}, we have that $A^\mu_{(0)}$ does not depend 
  on the coordinates spanning the boundary and then it can be brought outside the integral \eqref{sou_ter_tem}
  leading to
   \begin{equation}
    A_{(0)}^0 \int d^{d}x\  \rho\ .  
  \end{equation}
  The integral of the charge density over the boundary manifold gives the total charge. Hence $A_{(0)}^0$
  is naturally identified with the chemical potential $\mu$,
   \begin{equation}\label{chem_den}
    \mu \int d^{d}x\  \rho\ .  
  \end{equation}
  Being $\mu$ a source, we can compute the one-point function of the conjugate operator by means of functional differentiation
  of the partition function with respect to $\mu$ itself. Doing so, from \eqref{chem_den} we see that we get the expectation 
  value the charge density. If we follow the holographic prescription, we actually compute the one-point function of the charge density by means 
  of the functional differentiation of the dual on-shell gravity action. Proceeding in this way, we discover that the 
  one-point function of the charge density coincides with the subleading term in the UV expansion of the the time component of the bulk field. 
  Hence the notation introduced in \eqref{asyphi} is physically motivated.

   \vspace{20pt}
  \noindent
  {\bf Spontaneous condensation:} As described in the case of the vector field, also in relation to the bulk scalar we need to define 
  a well-posed variational problem. This can be achieved considering boundary conditions which either
  fix the value of the leading term or the subleading term on \eqref{asypsi}. According to the choice we make,
  we interpret the fall-off which has been fixed by the boundary condition as the source. Namely, if we fix $\psi_1$,
  then $\psi_2$ will be interpreted as the conjugate operator and vice-versa \cite{Klebanov:1999tb}.
  The choice of bulk boundary conditions is sometimes referred to, in a bulk perspective, as choice of ``quantization'' scheme.
  In other terms, it corresponds to the choice of the space of quantum states of the bulk model.
  
  Having chosen either bulk ``quantization'', we are interested in studying circumstances where the scalar operator
  acquires an unsourced, non-trivial VEV. Indeed we seek for spontaneous scalar condensation, namely a spontaneous 
  symmetry breaking leading to the superconducting phase. As a consequence we will always consider a scalar vanishing source.
  To comply with the existing literature we adopt the following definitions for the condensates in either quantization scheme
  \begin{eqnarray}\label{quantiquanti}
   &\psi_1 = 0\ , & \ \ \ \ \mathcal{h} O_2 \mathcal{i} = \sqrt{2}\ \psi_2\\
   &\psi_2 = 0\ , & \ \ \ \ \mathcal{h} O_1 \mathcal{i} = \sqrt{2}\ \psi_1\ .
  \end{eqnarray}

  \vspace{20pt}
  \noindent
  {\bfseries Scalings:} A direct study of the equations of motion shows that the system 
  is invariant under the following scalings:
  \begin{equation}\label{scachi}
   e^\chi \rightarrow a^2 e^\chi\ , \ \ \ \
   t \rightarrow a t \ , \ \ \ \
   \phi \rightarrow \phi/a\ ,
  \end{equation}
  \begin{equation}
   r \rightarrow a r\ , \ \ \ \
   t \rightarrow a t\ , \ \ \ \
   L \rightarrow a L\ , \ \ \ \
   q \rightarrow q/a\ ,
  \end{equation}
  and
  \begin{equation}\label{adssca}
   r \rightarrow a r\ , \ \ \ \
   (t,x,y) \rightarrow (t,x,y)/a \ , \ \ \ \
   g \rightarrow a^2 g\ , \ \ \ \
   \phi \rightarrow a \phi\ .
  \end{equation}
  The former scaling can be employed to put $\chi = 0$ at the boundary and so
  it can be used to impose \emph{a posteriori} asymptotic $AdS$-ness to our 
  solutions%
  \footnote{As a technical note, the possibility of exploiting the scaling \eqref{scachi}
  to fix $\chi=0$ at the boundary proves to be very convenient when writing the numerical 
  code. Indeed, we can solve leaving $\chi$ generic at the boundary and then obtain easily 
  an asymptotically $AdS$-solution.}. 
  The latter two scalings, instead, can be used to to put the $AdS$ radius
  $L$ and the horizon radius $r_h$ to $1$  without spoiling the generality of the treatment.

  The scaling \eqref{adssca} is the bulk manner to encode the scaling invariance 
  of the dynamics of theory living at the boundary. As a consequence of the scaling,
  only ratios of physical quantities have usually a relevant physical meaning, e.g. $T/\mu$.
  Indeed, for instance, the neutral black-hole (which is an $AdS$-Schwarzschild configuration) 
  is dual to a CFT put at finite temperature where all the finite values of the temperature are 
  physically equivalent. They in fact correspond to bulk solutions related by the scaling \eqref{adssca}.  
  To work with scale invariant quantities we need to improve the definition of the condensates 
  given in \eqref{quantiquanti}, namely we introduce
  \begin{eqnarray}\label{scaconde}
   \mathcal{h} \tilde{O}_2 \mathcal{i} = \frac{\mathcal{h} O_2 \mathcal{i}}{\mu^2}\ , \ \ \ \
   \mathcal{h} \tilde{O}_1 \mathcal{i} = \frac{\mathcal{h} O_1 \mathcal{i}}{|\mu|}\ .
  \end{eqnarray}
  Actually we have considered dimension-less ratios. Referring instead to a canonical treatment we should define
   \begin{eqnarray}
   \mathcal{h} \tilde{O}_2 \mathcal{i}_{\text{canonical}} = \frac{\mathcal{h} O_2 \mathcal{i}}{|\rho|}\ , \ \ \ \
   \mathcal{h} \tilde{O}_1 \mathcal{i}_{\text{canonical}} = \frac{\mathcal{h} O_1 \mathcal{i}}{\sqrt{|\rho|}}\ .
  \end{eqnarray}
In a similar fashion we also define the scale-invariant temperature
\begin{equation}
 \tilde{T}= T /\mu\ .
\end{equation}

%

\vspace{20pt}
\noindent
{\bf The normal phase:} In the superconductor jargon, the normal phase is where the system is not superconducting
and there is no condensate. This maps to a hairless black hole solution which is then 
a charged Reissner-Nordstr\"{o}m black hole. In terms of the ansatz \eqref{gans} is given by
\begin{eqnarray}
 &\chi(r)& = 0\\
 & g(r)  &= r^2 \left(1 - \frac{r_h^3}{r^3} \right) + \frac{\mu^2 r_h^2}{4 r^2} \left( 1 - \frac{r}{r_h} \right)\ .
\end{eqnarray}
The gauge field is instead
\begin{equation}\label{gau_np}
 A_t = \phi = \mu \left(1 - \frac{r_h}{r}\right)\ ,
\end{equation}
and, recalling \eqref{asyphi}, we have $\rho = \mu r_h$ and all the other gauge field components are vanishing.

In the extremal (i.e. $T=0$) circumstance, the expression 
for the temperature \eqref{temp} yields the extremal value of the 
chemical potential. Indeed, putting $\psi = 0$ and $\chi = 0$ (as we are in the normal phase)
into \eqref{temp} and asking $T=0$ we obtain
\begin{equation}
 \phi'(r_h) = \frac{2 \sqrt{3}}{L}\ .
\end{equation}
Comparing with the derivative of the gauge profile in the normal phase \eqref{gau_np},
we have
\begin{equation}\label{extremal_gau}
 \mu _{(T=0)} = \frac{2\sqrt{3}}{L}\, r_h\ , \ \ \ \
 \phi_{(T=0,\text{n.p.})} = \frac{2\sqrt{3}}{L}\, r_h  \left(1 - \frac{r_h}{r}\right)\ .
\end{equation}

\vspace{20pt}
\noindent
{\bf Near-horizon geometry:} Let us study the near-horizon geometry of the Reissner-Nordstr\"{o}m black
hole solution in $d$ spatial dimensions (the action \eqref{actia} corresponds
to $d=2$). In the zero-temperature limit, the horizon radius is given by
\begin{equation}
 r_H^2=\frac{1}{2d}\frac{(d-2)^2L^2\mu^2}{(d-1)}\, .
\end{equation}
To obtain the near-horizon behavior of the metric we expand the function $g(r)$ around $r_h$%
\footnote{The function $g(r)$, sometimes called \emph{blackening factor} and it has been introduced in \eqref{gans}.}
\begin{equation}
g(r) = g(r_h)+g^\prime(r_h)\tilde{r}+\frac{1}{2}g^{\prime\prime}(r_h)\tilde{r}^2 + ... \,,
\end{equation}
where $r=r_h+\tilde{r}$ and $\tilde{r}\rightarrow0$.
For the Reissner-Nordstr\"{o}m black hole we find
\begin{equation}\label{pappetta}
g(r_h)=0\,, \qquad g^\prime(r_h)\propto T=0\,,\qquad g^{\prime\prime}(r_h)=\frac{2d(d-1)}{L^2}\, ,
\end{equation}
hence
\begin{equation}
g(\tilde{r}) =  \frac{d(d-1)}{L^2}\, \tilde{r}^2 + ...\,,
\end{equation}
Notice that in this case the first derivative of the metric function $g$ at the horizon \eqref{pappetta} is proportional 
to the temperature. This result agrees with what we have obtained in Subsection \ref{stu}%
\footnote{Indeed, referring to equation \eqref{gen_sha}, we have $g = a b$. 
For the RN solution we have $a(r)=1$ (i.e. $\chi = 0$ referring to the notation of \eqref{gans}).
Hence, here we have $g=b$. Putting these things together and comparing with formula \eqref{temp}
we have $g'(r_H) \propto T$.}.
As a consequence, the near-horizon metric is
\begin{equation}\label{nh_t0}
ds^2_{(T=0,\mbox{\scriptsize{n.-h.}})}\simeq -d(d-1)\frac{\tilde{r}^2}{L^2}dt^2+\frac{r_H^2}{L^2}d\vec{x}^2+\frac{L^2}{d(d-1)\tilde{r}^2}d\tilde{r}^2\,,
\end{equation}
from which we recognize the $AdS_2\times R^{d-1}$ form. 
In particular, the $AdS_2$ radius squared is 
\begin{equation}
 L_{(2)}^2=\frac{L^2}{d(d-1)} \ .
\end{equation}

In \cite{Faulkner:2009wj} the near-horizon $AdS_2 \times R^{d-1}$, which is generic for
finite-temperature and finite-charge-density solutions, has been interpreted as the dual
of an IR quantum critical dynamics%
\footnote{See also \cite{Faulkner:2010da}. To have a top-down perspective on IR $AdS_2$ geometries 
we refer to \cite{Donos:2012yi}.}. 
More precisely, the $AdS_2$ geometry should correspond to 
a $(0+1)$-dimensional CFT describing \emph{local} quantum critical behavior, where time 
(and therefore energy as well) can be rescaled independently of space (or momentum).
The possibility of scaling the energy of a fermionic operator independently of the momentum
is an intuitive hint that local quantum theories can share some phenomenological features 
with systems having a Fermi surface (see for instance \cite{Liu:2009dm}). Actually, excitations close to the Fermi surface are
vanishing energy and finite momentum states. We refer to \cite{Faulkner:2009wj} and related 
literature for further detail. Nevertheless let us stress that the local IR critical behavior 
is an emergent phenomenon in the same sense as the emergent degrees of freedom we referred to in Section \ref{wout}
which arise from many-body collective dynamics. As a final comment, let us underline that quantum states
featuring an emergent time scaling invariance are similar to the marginal Fermi-liquid advanced 
(in a non-holographic context) to account for strange metal behavior in copper oxides systems \cite{Varma:1989zz}.

\vspace{20pt}
\noindent
{\bf Zero-temperature, near-horizon scalar equation:} As we will see in Subsection \eqref{scastab}, the study of the near-horizon limit
of the scalar equation at vanishing temperature is relevant for the study of the 
low-temperature stability of the Reissner-Nordstr\"{o}m black hole (i.e. the stability
of the normal phase at low $T$).

We recall the $T=0$ near-horizon form of the metric given in \eqref{nh_t0}
and the expression for the gauge field at extremality \eqref{extremal_gau}.
In the $T=0$ near-horizon limit, the scalar equation \eqref{eqscal} assumes the following form%
\footnote{Being in the normal phase, we have $\chi=0$. Furthermore, in the coefficient of the 
$\psi'$ term we can neglect the $2/r$ which remains finite at the horizon with respect to 
$g'/g$ which explodes.}
\begin{equation}
 \psi'' + \frac{2}{\tilde{r}} \psi' + \frac{2 q^2 - m^2}{d(d-1)} \frac{1}{\tilde{r}^2} \psi = 0\ .
\end{equation}
This equation is equivalent to a free scalar on $AdS_2$ possessing effective mass
\begin{equation}\label{eff_mass}
 m^2_{(\text{eff})} = \frac{2 q^2 - m^2}{d(d-1)} \ .
\end{equation}

\subsection{Stability}
\label{scastab}

Let us still consider explicitly the bulk Einstein-Maxwell-Higgs model in $d+1$ dimensions 
in the presence of a negative cosmological constant (namely the generalization of \eqref{actia}
to $d+1$ dimensions). When the gauge and scalar fields are vanishing, 
we have the usual $AdS_{d+1}$ solution. We define the coordinate $z=r_h/r$ where $r$ is
the standard radial coordinate and we set the horizon radius $r_h$ to one as 
described in Subsection \eqref{stu}, so
\begin{equation}
 z = \frac{1}{r}\ .
\end{equation}
The conformal boundary corresponds to $z=0$ and the black hole horizon is at $z=1$.

We assume a near-boundary behavior for the scalar field of the power-law type. 
From the term-wise study of the near-boundary expansion of the scalar field equation in $d+1$ dimensions we have
that $\psi$ behaves as
\begin{equation}\label{UVfo}
 \psi \sim \psi_{(-)} z^{\Delta_-} + \psi_{(+)} z^{\Delta_+}\ .
\end{equation}
The exponents $\Delta_\pm$ are related to the mass of the scalar and the dimensionality
of the space-time in the following fashion
\begin{equation}\label{confma}
 \Delta (\Delta - d) = m^2 L^2\ ,
\end{equation}
where $L$ is the $AdS_{d+1}$ curvature radius.
Solving \eqref{confma} we get
\begin{equation}
 \Delta_{\pm} = \frac{d \pm \sqrt{d^2 + 4m^2 L^2}}{2}\ .
\end{equation}
Requiring real solutions for $\Delta_{\pm}$, we have
\begin{equation}\label{BF_bound}
 m^2 > - \frac{d^2}{4 L^2}\ .
\end{equation}
From a boundary field theory viewpoint, this requirement corresponds to
the unitarity bound \cite{Klebanov:1999tb}. From the bulk perspective,
requiring the reality of the solutions for $\Delta_{\pm}$ amounts to require 
a stable bulk geometry with respect to the scalar fluctuations. This is known in the
literature as the Breitenlohner-Friedmann bound. Note that, given the curved character 
of the $AdS$ geometry, negative values of the scalar mass (if not too negative in the sense 
of the BF bound \eqref{BF_bound}) do not lead to an instability.

There is a range of values for the squared mass
for which both the UV fall-off's \eqref{UVfo} of the scalar field are normalizable.
Here with ``normalizability'' we mean the finiteness of the integral of the field Lagrangian density
on the surface defined by constant radial coordinate $r_s$ in the limit $r_s\rightarrow \infty$. 
Let us have a sketchy look at it:
consider a $d+1$-dimensional bulk and the kinetic term of the scalar field.
In the large $r$ limit, on a fixed radius shell, the integrand goes as
\begin{equation}\label{normalizz}
 r^d (\partial \psi)^2 \sim r^{d-2\Delta-2}
\end{equation}
where the factor $r^d$ is given by the volume of the hyper-surface at fixed radius.
In order to have a normalizable fall-off we need to require
\begin{equation}
 \Delta > \frac{d}{2 L} - 1 \Rightarrow m^2 < - \frac{d^2}{4 L^2} + 1
\end{equation}
Packing together the BF bound and the requirement that both fall-off's 
of the scalar are normalizable we obtain the following range for the scalar square mass
\begin{equation}
 -\frac{d^2}{4 L^2} < m^2 < -\frac{d^2}{4 L^2} + 1\ .
\end{equation}

\subsection{IR instability and hair formation}

The Asymptotically Anti-de Sitter ($AAdsS$) Reissner-Nordst\"{o}m black hole solution, is not thermodynamically 
favored at very low temperature. This means that if we start from an $AAdsS$ Reissner-Nordst\"{o}m solution
and we progressively lower the temperature we encounter an instability. Specifically, the black hole
becomes instable towards the formation of scalar hair and it is by studying the scalar fluctuations that one
can characterize the instability. 

The Breitenlohner-Freedman bound \eqref{BF_bound} was referring to an $AdS_5$ geometry,
which corresponds to the large radius asymptotic geometry of the solution in exam. However the instability 
towards hair formation emerges in the near-horizon region. As observed in Subsection \ref{stu},
the near-horizon geometry contains an $AdS_2$ factor
whose curvature radius is $L_{(2)}=L/6$.
Also this IR $AdS_2$ geometry has its own BF bound. As we have seen, still referring to the 
near horizon region, the scalar field equation reduces to a free scalar field on $AdS_2$ 
characterized by an effective value of the scalar mass given in \eqref{eff_mass}.
To judge the IR stability of the $T=0$ Reissner-Nordstr\"{o}m solution
we have to study the BF bound for the near-horizon effective scalar mass, namely
\begin{equation}
 L^2_{(2)} m_{(\text{eff})}^2 < -\frac{1}{4}\ ,
\end{equation}
which, in terms of the UV radius of curvature becomes
\begin{equation}
 \frac{L^2}{6} m_{(\text{eff})}^2 < -\frac{1}{4}\ .
\end{equation}
Notice that here we are demanding for the possibility of having an instability.
Indeed we intend to consider systems which develop charged scalar hair in the bulk,
this corresponding to a superconducting condensation from the boundary system perspective.
As argued in Subsection \ref{stu}, we exploit a scaling invariance of the equations
of motion to fix from now on $L=1$. We have the following explicit bound on the effective mass
\begin{equation}
 m_{(\text{eff})}^2 < -\frac{3}{2}\ ,
\end{equation}
which translates into a bound on the UV mass of the scalar
\begin{equation}
 m^2 < 2q^2 - \frac{3}{2}\ .
\end{equation}
All in all, to have a UV asymptotic $AdS_5$ geometry and an instable
IR $AdS_2$ factor at $T=0$, we require the ``bare'' mass of the scalar to fall
in the following interval
\begin{equation}
 -\frac{9}{4} < m^2 < 2 q^2 - \frac{3}{2}\ .
\end{equation}

\subsection{On-shell action for the hairy black hole and the free energy}

In its strongest version, a holographic correspondence conjectures the identity of the 
partition functions of the two dual theories. We adopt a semi-classical approximation 
for the weakly coupled, low-energy gravity model. The partition function is then approximated 
with the exponential of the on-shell action. Enforcing the duality, this is read as the generating functional 
of the conformal field theory. Considering the Euclidean framework on both sides of the duality,
we have that the gravity Euclidean on-shell action corresponds, on the field theory side, 
to the thermodynamical potential whose exponential gives the generating functional, namely
the free energy of the boundary theory. 

On a computational level, to study the free energy of the field theory we then need to
consider the Euclidean version of the gravity action \eqref{actia},
\begin{equation}\label{aceu}
 S^{(\text{Eu})} = - \int d^4 x \sqrt{-g}\, {\cal L}\ .
\end{equation}
Working with the ansatz specified in \eqref{phipsians} and \eqref{gans},
we have neither dependence on the coordinate $x$ nor on the $A_x$ gauge field component. 
As a consequence, in the right-hand-side of the $xx$ 
Einstein equation \eqref{Ei}, only the terms proportional to $g_{xx}$
do remain. We have then
\begin{equation}
 G_{xx} = \frac{1}{2} r^2 ({\cal L} - R)\ .
\end{equation}
An identical reasoning holds for the $yy$ component as well.

Let us perform some manipulations to get a nice and useful expression for the Euclidean
on-shell action. Recall the definition of the Einstein tensor 
\begin{equation}\label{simsim}
 G_{\mu\nu} = R_{\mu\nu} - \frac{R}{2} g_{\mu\nu}\ ;
\end{equation}
taking the trace of \eqref{simsim} we get
\begin{equation}\label{tracia}
 G^\mu_{\ \mu} = - R\ .
\end{equation}
In addition, combining \eqref{tracia} with equation \eqref{simsim}, we obtain
\begin{equation}
 - R = G^\mu_{\ \mu} = G^t_{\ t} + G^r_{\ r} + {\cal L} - R\ .
\end{equation}
We can therefore express the Lagrangian density in a very useful fashion, namely
\begin{equation}\label{call}
 {\cal L} = - G^t_{\ t} - G^r_{\ r} = -\frac{1}{r^2} [(rg)' + (rge^{-\chi})'e^\chi]\ ,
\end{equation}
where at last we have expressed the result in terms of the functions appearing in the ansatz
\eqref{gans}. In terms of these functions, the metric determinant is expressed as follows
\begin{equation}\label{deto}
 \sqrt{-g} = e^{-\chi} r^2\ .
\end{equation}
Plugging \eqref{deto} and \eqref{call} into \eqref{aceu}
allows us to write explicitly the on-shell value of the Euclidean action, namely
\begin{equation}\label{seu}
 S^{(\text{Eu})}_{(\text{on-shell})} = \int d^3x\ \int_{r_h}^{r_\infty} dr [2r g e^{-\chi/2}]'\ .
\end{equation}
Recall indeed that the arguments leading to \eqref{call} relied on the form of 
the ansatz which, once the equations of motion for $g$ and $\chi$ have been solved, 
represents a vacuum solution for the action \eqref{aceu}.
The integrand appearing in \eqref{seu} has the nice feature of being expressed as a
radial total derivative. The on-shell action reduces therefore to boundary terms.
Moreover, the vanishing of the function $g$ at the horizon implies that the only 
non-null contribution to the on-shell action comes from the conformal 
boundary, so
\begin{equation}\label{osb}
 S^{(\text{Eu})}_{(\text{on-shell})} = \left. \int d^3x\ 2r\ ge^{-\chi/2} \right|_{r=r_\infty}\ .
\end{equation}

Since we are interested exclusively in spontaneous (i.e. unsourced) scalars, we have always
either vanishing $\psi_1$ or $\psi_2$ in the near-boundary scalar expansion \eqref{asypsi}.
In such circumstances, an asymptotic study of the Einstein equations reveals that
\begin{equation}\label{g_chi}
 e^{-\chi} g = \frac{r^2}{L^2} + \frac{C_{(-1)}}{r} + ...
\end{equation}
where the dots stand for subleading terms in the large $r$ limit.
The asymptotic behavior of \eqref{g_chi} reveals that the on-shell action \eqref{osb}
is divergent. To address correctly the on-shell action divergence we must introduce appropriate
counter-terms as described in \cite{Hartnoll:2008kx} and references therein. We recall here the 
necessary counter-terms and, skipping the intermediate passages, we give an expression for the renormalized 
on-shell action. Relying still on the spontaneity requirement, the only non-vanishing counter-term 
we need is the so-called Gibbons-Hawking-York term \cite{York:1972sj,Gibbons:1976ue} plus a boundary cosmological constant, namely
\begin{equation}\label{GHY}
 S_{\text{G.H.Y.}} = \left. \int d^3 x\ \sqrt{-\gamma_\infty} \left(- 2K + \frac{4}{L}\right) \right|_{r=r_\infty}\ ,
\end{equation}
where $r_\infty$ is the radial value identifying the radial regularizing shell ($r_\infty$ is eventually to 
be sent to infinity and its finiteness corresponds to the presence of a non-vanishing large $r$ cut-off).
The metric $\gamma_\infty$ is that induced by the bulk metric on the radial shell at $r=r_\infty$ and $K$
represents the extrinsic curvature, still referred to the radial shell at $r_\infty$.

A similar holographic renormalization procedure will be analyzed (there in full detail) while
treating the bulk fluctuations. For the moment being, let us simply report that when $r_\infty$
is actually sent to $\infty$ the regularized action 
\begin{equation}
 S_{(\text{reg})} = S^{(\text{Eu})}_{(\text{on-shell})} +  S_{\text{G.H.Y.}}\ ,
\end{equation}
gives the renormalized action
\begin{equation}
 S_{(\text{REN})} = \int d^3 x\ C_{(-1)}\ ,
\end{equation}
where $C_{(-1)}$ has been introduced in the relation \eqref{g_chi}
giving the near-boundary expansion of $e^{-\chi}g$. 

The free energy is given by $F = - T\ln(Z)$ where $Z$ is the partition function. Exploiting the gauge/gravity correspondence
we have that the field theory partition function coincides with that of the gravity model. In the semi-classical approximation
we are adopting, this latter is given by $Z = e^{-S}$. All in all, we have
\begin{equation}\label{freen}
 F = T S_{(\text{REN})} = T \int d^3 x\ C_{(-1)} = C_{(-1)} V_2 =  - \left( r_h^3 - \frac{1}{4} \mu \rho \right) V_2\ ,
\end{equation}
where $V_2$ indicates the volume of the 2-dimensional spatial boundary manifold. In the passages we have used the fact that the 
integral along the compact, Euclidean time direction gives the inverse temperature and considered the normal phase 
solution specified in \eqref{gau_np}.

\section{Equilibrium}

An equilibrium solution of the boundary theory corresponds to a vacuum solution of the dual gravity model.
In the previous sections we have introduced all the necessary ingredients to study the boundary thermal system
in terms of the thermodynamical physical quantities (e.g. $T$, $\mu$, $\rho$,...). We here report the results of the numerical analysis.

\subsection{Condensation}

Considering the system at progressively lower values of the scale-invariant parameter $\xi=T/\mu$,
we encounter the possibility of a phase transition. This means that, below a critical value $\xi_c$,
the gravitational system admits two different solutions: besides the Reissner-Nordstr\"{o}m black hole
we can have a hairy black hole solution. The actual occurrence of the phase transition from the normal to a condensed 
phase is understood in terms of the free energy. The solution corresponding to a lower value of the free energy is 
thermodynamically favored.

\begin{figure}[ht]
\centering
\includegraphics[width=70mm]{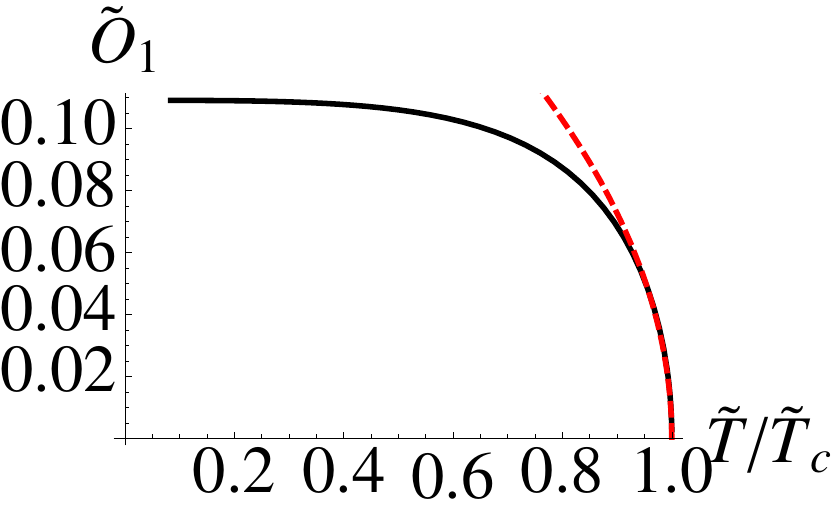} 
\includegraphics[width=70mm]{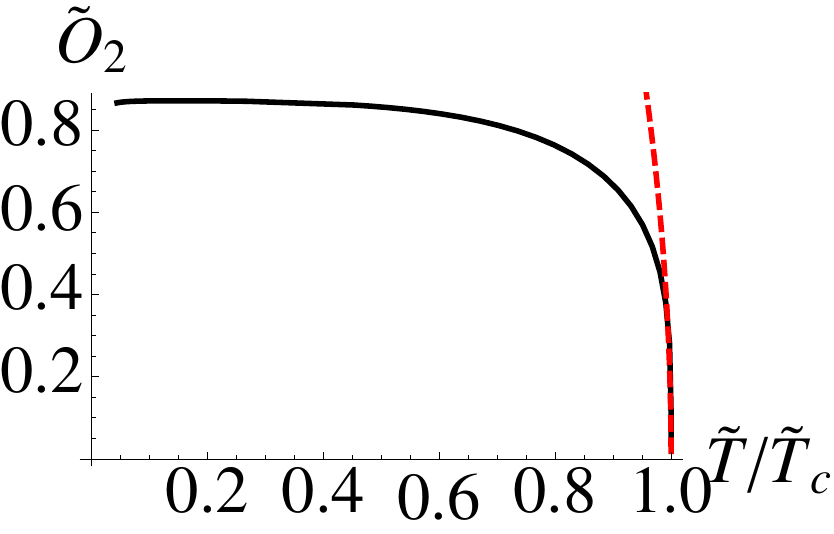} 
\caption{The two plots show the scalar condensates corresponding to the two $AdS$ quantization 
schemes with solid lines. The dashed lines represent instead the square root behavior expected 
from mean-field theory.}
\label{condi}
\end{figure}

The hairy black hole solution has a non-trivial scalar profile. This is dual to 
the condensation of a scalar charged operator in the boundary model. 
As we have already stated, we consider spontaneous condensation where the condensate 
is unsourced. In relation to the two possible quantizations we define the condensates
as in Equation \eqref{quantiquanti} and consider their scale invariant version \eqref{scaconde}.
In Figure \ref{condi} we report the results of the numerical analysis. As the temperature 
is lowered below a critical value the condensate operator acquires a non-vanishing VEV which is plotted.

Let us rely on some salient features of the plots. A first technical observation is 
that the flat behavior at low-temperature is a good feature of a backreacted treatment. In fact, 
as we have already argued, a probe analysis would have been unreliable at low-temperature.
The results referring to the two condensates (obtained with the two quantization schemes) are qualitatively 
similar. In particular (as first observed in \cite{Hartnoll:2008kx}) they show mean-field behavior close to $T=T_c$, namely%
\footnote{A discussion of a zero-temperature system featuring non-mean-field scalings is 
given in \cite{Evans:2010np}.}
\begin{equation}\label{campomedio}
 \mathcal{h} O_i\mathcal{i} \sim \left(1 - \frac{T}{T_c}\right)^{1/2}\ .
\end{equation}
Such a mean-field behavior is pictorially highlighted by the red dashed lines which 
precisely correspond to the right-hand-side of \eqref{campomedio}.

\section{Transport}
\label{trans}

Till now we have considered only the equilibrium properties of the holographic superconductor \eqref{actia}.
Now we turn the attention on some of its (slightly) out-of-equilibrium features and so to its transport properties.
The study of the transport of the system consists in the analysis of the linear response of the system itself to fluctuations of the external sources.
The legitimacy of the linear response approach needs the source variations to be in some sense ``small''.
More quantitatively, the higher order terms in the fluctuations must be negligible when compared with the linear order.
In the action of the boundary theory, an external source is introduced with a generic term as follows 
\begin{equation}
 \int d^Dx\ J^\mu A_\mu^{(0)} \ ,
\end{equation}
where $D$ is the dimensionality of the boundary manifold. The space-time label $\mu$ goes
over $\{0,1,..,D-1\}$ and $J^\mu$ represents the boundary operator corresponding to the current
sourced by $A_\mu^{(0)}$. 
The holographic prescription (also called \emph{holographic dictionary}) 
put into correspondence the source term $A_\mu^{(0)}$ to the boundary value of the
associated bulk gauge field $A_\mu$.
So, a small variation of the source is associated, from the dual gravitational standpoint, to a small
fluctuation of the corresponding bulk dynamical field boundary condition.
Recall that at the classical level, studying the bulk system and considering small boundary fluctuations 
means that we are interested in the analysis of the fluctuations of the gravitational bulk fields about their background values.
This fluctuation analysis about the background is done up to the linear order in the fluctuation fields.

\subsection{Bulk fluctuations}

We are interested in studying the transport due to the vector mode
fluctuations along (say) the spatial direction $x$. We will stick to 
space-independent fluctuations (or, equivalently, fluctuations at zero
momentum $k$). In such a circumstance, the vector 
fluctuations of the bulk system involve just the fluctuations 
for the gauge field component $A_x$ and the metric component $g_{tx}$.

The smallness of the fluctuations justifies the linear approximation, meaning that we 
consider the bulk equations at linear order in the fluctuations.
Observe that this, at the level of the action, is equivalent to retain only
the quadratic terms.
More specifically, we consider space-independent fluctuations whose 
time dependence is harmonic, i.e. $e^{-i\omega t}$. For a study of the 
momentum-dependent fluctuations and the hydrodynamic behavior of the 
holographic superconductor we refer to \cite{Amado:2009ts}.

In order to analyze the $g_{tx}$ metric fluctuations we ought to study the
$tx$ Einstein equation. We find the following first order equation
which couples the metric vector mode $g_{tx}$ with the fluctuations 
of the gauge field along $x$, namely
\begin{equation}\label{metflu}
 g'_{tx} - \frac{2}{r} g_{tx} + \phi' A_x = 0\ .
\end{equation}
We remind the reader that $\phi' = E$ represents the radial electric field
in the bulk. The fact that the vector fluctuations of the metric are governed 
by a simple first order differential equation is a special consequence of the 
fact that we are limiting ourselves to the zero-momentum case.

Switching our attention to the Maxwell equation along $x$ for the fluctuations of the gauge field,
we have
\begin{equation}\label{vecflu}
 A_x'' + \left(\frac{g'}{g} - \frac{\chi'}{2}\right) A_x'
 + \left( \frac{\omega^2}{g^2}e^\chi - \frac{2 q^2\psi^2}{g}\right) A_x = 
 \frac{\phi'}{g}e^\chi \left(-g_tx'+\frac{2}{r}g_{tx}\right)\ .
\end{equation}
In order to gain some insight into this equation, let us consider it first in the 
probe approximation. Recalling that the probe limit consists in sending the charge 
$q$ to infinity while keeping the product $q \phi$ finite, we see from the metric 
fluctuation equation \eqref{metflu} that the term in $\phi$ field drops away.
The dynamics of the metric fluctuations becomes independent and from the study 
of their equation we find either the trivial solution $g_{tx}=0$ or $g_{tx} \sim r^2$.
The latter, being diverging at the boundary, has to be discarded. The outcome of this
reasoning is that, in order to consider the vector fluctuations in the probe approximation
we can simply neglect the metric fluctuations and pose $g_{tx}$ to zero into 
Equation \eqref{vecflu}.

If we further neglect, just for one moment, the radial dependence of the vector fluctuations
and stick to the probe approximation, we obtain the following approximate equation for the 
gauge field fluctuations
\begin{equation}
 A_x''  + \left( \frac{\omega^2}{g^2}e^\chi - \frac{2 q^2\psi^2}{g}\right) A_x \sim 0\ .
\end{equation}
Such crude approximation makes it manifest that we are dealing with a sort of Brout-Englert-Higgs mechanism for the 
bulk gauge field; indeed, we see that the field $A_x$ acquires a mass related to the scalar field $\psi$.
In other words, whenever we are in the broken phase (i.e. when $\psi$ in non-trivial), the gauge 
field fluctuations are massive.

Going back to the fully backreacted system of equations for the fluctuations \eqref{metflu} and
\eqref{vecflu}, it is worthwhile to observe that we can derive an equation governing 
the vector field fluctuations where the metric fluctuations do not appear manifestly. 
Indeed, substituting \eqref{metflu} into \eqref{vecflu}, we have
\begin{equation}\label{traflu}
 A_x''+ \left(\frac{g'}{g} - \frac{\chi'}{2}\right) A_x'
 + \left[\left(\frac{\omega^2}{g^2}-\frac{\phi'^2}{g}\right)e^\chi
 -\frac{2q^2\psi^2}{g}\right]A_x = 0\ .
\end{equation}
On a practical level, the absence of the metric fluctuations $g_{tx}$ in Equation \eqref{traflu}
implies that we have the possibility of studying the fully backreacted 
electro-magnetic linear response of the system without having to deal with the geometry fluctuations. 
Note however that, substituting the equation for $g_{tx}$ into \eqref{vecflu},
we obtained a new contribution to the effective mass of the gauge field, namely $-\phi'^2/g$.

As observed in \cite{Hartnoll:2008kx} (from which all the present analysis is taken),
the possibility of having an equation for $A_x$ which does not depend explicitly 
on $g_{tx}$ has an important physical meaning. Being decoupled from $g_{tx}$,
the equations for the gauge fields are sensitive to a completely translation invariant
context. As we will shortly see in Section \ref{autsuc}, this has important consequences 
on the electric transport of the system and, specifically, it will lead to a divergent
D.C. conductivity also in the normal phase (not to be confused with the actual 
superconductivity occurring because of the presence of charged condensate in the 
broken phase).

\subsubsection{Ingoing IR boundary conditions}
\label{InOut}

The explicit fluctuation equations just obtained assumed harmonic time dependence.
In the linearized Maxwell equation the derivative with respect to time appears 
quadratically and actually Equation \eqref{vecflu} is insensitive to the sign of $\omega$. 
In addition, the oscillating factor $e^{-i \omega t}$ can be collected outside
and hence dropped. Instead, the linearized Einstein equation for the metric fluctuations
\eqref{metflu} does not contain time derivatives.
Therefore we have that, eventually, the fluctuation equation \eqref{traflu} obtained 
putting together \eqref{vecflu} and \eqref{metflu} is still insensitive to the 
sign of the frequency $\omega$.

Consider the following solution ansatz for the fluctuation equation in the near-horizon region:
\begin{equation}\label{nearhorans}
 e^{-i \omega t}\,(1-z)^{-i \alpha \omega} \left[a_0 + a_1 (1-z) + ... \right]\ .
\end{equation}
As the equation is not sensitive to the sign of $\omega$ our analysis here goes through for either signs.
In order to appreciate in which way along the radial coordinate the wave \eqref{nearhorans} propagates,
we can rewrite \eqref{nearhorans} as follows
\begin{equation}
 \begin{split}\label{travel}
 e^{-i \omega t - i \omega \alpha \ln(1-z)}  \left[a_0 +... \right]
 \sim e^{-i \omega (t - \alpha z)}  \left[a_0 +... \right]\ ,
 \end{split}
\end{equation}
in which we have considered the near-horizon assumption $z \sim 1$.
From Equation \eqref{travel} we understand that the sign of $\alpha$ coincides then with the sign of the speed of the propagating wave
along the radial direction (parametrized by $z$).
Recall that the horizon is located at $z=1$ and the conformal boundary is instead at $z=0$;
hence increasing $z$ corresponds to moving towards the black hole. 
In other terms, a positive $\alpha$ corresponds to an in-going wave.

In holography, it is crucial to distinguish between in-going and out-going solutions.
In fact, the holographic prescription to compute the correlation functions for a 
Minkowskian boundary theory, associates respectively in-going and out-going bulk solutions
to retarded and advanced boundary correlators (see \cite{Son:2002sd}).

As a final observation, notice that, from the generic near-horizon ansatz given in \eqref{nearhorans}, 
we read that the modulus of the fluctuation solution tends to a constant value in the proximity 
of the horizon. Such modulus is related to the first coefficient in the near-horizon expansion,
namely $a_0$. In contrast, the phase diverges at the horizon. Let us underline however that from 
Equation \eqref{travel} the speed of the propagating wave is related to the exponent $\alpha$ and therefore,
approaching the horizon, it approaches a finite and constant value.

\subsubsection{Holographic renormalization for the fluctuations}

The quadratic on-shell action for the fluctuations reduces to the following boundary term
\begin{equation}\label{flu2}
 S^{(2)}_{\text{on-shell}} = \int d^3 x\ e^{\chi/2} \left[ 
 -\frac{g}{2}e^{-\chi} A_x A_x' - g_{tx} g_{tx}' + \frac{1}{2}\left(\frac{g'}{g}-\chi'\right) g^2_{tx} \right]_{r=r_\infty}\ .
\end{equation}
It should be observed that we obtained only contributions from the boundary corresponding to the upper radial limit $r_\infty$;
instead, the contribution coming from the horizon at $r=r_h$ vanishes
because both the metric function $g$ and the vector fluctuation component $g_{tx}$ are zero on the horizon. 
Moreover, recall that the fields are assumed to be null at large values of the spatial
and temporal coordinates.

Since the quadratic action $S^{(2)}_{\text{on-shell}}$ is divergent it requires a regularization procedure. 
However, the divergent terms are only due to the metric fluctuations while the term contributed 
by the vector is finite. Let us look at the divergence of the on-shell action in greater detail.
From the term-wise, near boundary study of the fluctuation equations we have the following
asymptotic behaviors for the fluctuation fields
\begin{equation}\label{UV_fluct}
 A_x = A_x^{(0)} + \frac{A_x^{(1)}}{r} + ...\ , \ \ \ \
 g_{tx} = r^2 g_{tx}^{(0)} + \frac{g_{tx}^{(1)}}{r} + ... \ .
\end{equation}
Henceforth we specialize the analysis to the case $\psi_1 = 0$, i.e. ``spontaneous condensation'' for $\psi_2$.
In such a framework, from Equation \eqref{eq_chi}, we have
\begin{equation}
 \chi' = 4 \frac{\psi_2^2}{r^5} + {\cal O}(r^{-6})\ ,
\end{equation}
hence
\begin{equation}
 e^\chi = 1 + {\cal O}(r^{-4})\ .
\end{equation}
We remind the reader that the background metric function $g$ 
\begin{equation}
 g = \frac{r^2}{L^2} - \frac{\epsilon L^2}{2 r} + ...\ ,
\end{equation}
from which we have
\begin{equation}
 \underbrace{e^{\chi/2}}_{1}\ \underbrace{g}_{r^2}\ \underbrace{e^{-\chi}}_{1}\ \underbrace{A_x}_{1}\ \underbrace{A_x'}_{r^{-2}} \sim 1\ .
\end{equation}
So, as anticipated, we do not need to regularize the piece in the action involving the gauge field. More precisely we have
\begin{equation}
 e^{\chi/2} \left( - \frac{g}{2} e^{-\chi} A_x A_x'\right)  = \frac{1}{2} \frac{1}{L^2} A_x^{(0)} A_x^{(1)} + ...\ .
\end{equation}
Conversely, the terms involving the metric fluctuations behave as follows as $r\rightarrow\infty$,
\begin{equation}
 \underbrace{e^{\chi/2}}_{1}\ \underbrace{g_{tx}}_{r^2}\ \underbrace{g_{tx}'}_{r}\ \sim r^3\ ,
\end{equation}
\begin{equation}
 \underbrace{e^{\chi/2}}_{1}\ \underbrace{(g'/g)}_{r^{-1}}\ \underbrace{g_{tx}^2}_{r^4} \sim r^3\ ,
\end{equation}
\begin{equation}
 \underbrace{e^{\chi/2}}_{1}\ \underbrace{\chi'}_{r^{-5}}\ \underbrace{g_{tx}^2}_{r^4} \sim 1/r\ .
\end{equation}
Keeping track also of the numerical coefficients we have
\begin{equation}
 - e^{\chi/2} g_{tx}\, g'_{tx} = - 2 r^3 g_{tx}^{(0)} g_{tx}^{(0)} - g_{tx}^{(0)} g_{tx}^{(1)} + ...\ ,
\end{equation}
and
\begin{equation}
  e^{\chi/2}\, \frac{1}{2}\, \frac{g'}{g}\, g_{tx}^2 = r^3 g_{tx}^{(0)}g_{tx}^{(0)} 
  + 2 g_{tx}^{(0)} g_{tx}^{(1)}
  + \frac{3}{4}\,\epsilon L^4\, g_{tx}^{(0)} g_{tx}^{(0)}
  + ...\ .
\end{equation}
Packaging everything together, we have the quadratic on-shell action for the fluctuations
\begin{equation}
 S^{(2)}_{\text{on-shell}} = \int d^3 x\ \left[
  \frac{1}{2} \frac{1}{L^2} A_x^{(0)} A_x^{(1)} 
  - r^3 g_{tx}^{(0)} g_{tx}^{(0)} 
  + g_{tx}^{(0)} g_{tx}^{(1)}    
  + \frac{3}{4}\,\epsilon L^4\, g_{tx}^{(0)} g_{tx}^{(0)}
  + ... \right]\ .
\end{equation}

To cure the divergence of $S^{(2)}_{\text{on-shell}}$ we apply the holographic renormalization procedure which amounts to
regularizing the action by means of counter-terms 
and than considering the limit $r_\infty \rightarrow \infty$ for the regularized action.
The standard procedure to implement the holographic renormalization requires the introduction of the following counter-term
(see \cite{Hartnoll:2008kx,Liu:1998bu} and references therein):

\begin{equation}\label{cgK}
 S_{C.T.} = \int d^3x\, \left. \sqrt{-\tilde{g}}\ \left(-2 K + \frac{4}{L}\right) \right|_{r=r_\infty} \ .
\end{equation}
This is the same term \eqref{GHY} which we have already encountered studying the background action.
Note that here $\tilde{g}$ is the metric induced on the $3$-surface defined by a constant radius and
the last term inside the brackets in \eqref{cgK} is a boundary cosmological constant.

The quadratic regularized action for the fluctuations of the fields $A_x$ and $g_{tx}$ is 
\begin{equation}
 S^{(2)}_{\text{reg}} = S^{(2)}_{\text{on-shell}} + S_{C.T.}  \ .
\end{equation}
As anticipated, we would like to study the terms of $S_{reg}$ which, in the limit $r\rightarrow \infty$, 
are quadratic in the fluctuations,
\begin{equation}\label{renofluc2}
 S^{(2)}_{\text{reg}} = \left.\lim_{r_\infty\rightarrow \infty} S_{\text{reg}} \right|_{{\cal O}(2)_{A_x,g_{tx}}}\ .
\end{equation}
Actually, as far as the transport is concerned, we need to renormalize the quadratic on-shell
action for the fluctuations \eqref{flu2} according to \eqref{renofluc2}, as we now do explicitly.

Let us start considering the expansion of the determinant factor appearing in \eqref{cgK} for large values 
of the radial coordinate,
\begin{equation}
 \sqrt{-\tilde{g}} = \frac{r^3}{L} - \frac{\epsilon L^3}{4} 
                     + \frac{1}{2}\, L r^3 \, g_{tx}^{(0)}g_{tx}^{(0)} + L\,  g_{tx}^{(0)}g_{tx}^{(1)} 
                     +\frac{1}{8} \,L^5 \epsilon\, g_{tx}^{(0)}g_{tx}^{(0)}+ ...\ .
\end{equation}
The metric induced on a radial shell is given by
\begin{equation}\label{boumet}
 d\tilde{s}^2 = -g\, e^{-\chi} dt^2 + r^2 \left(dx^2+dy^2\right) + g_{tx} \left(dx\, dt + dt\, dx\right)\ ,
\end{equation}
whereas the extrinsic curvature $K$ of a constant radius surface is given by
\begin{equation}\label{extrins}
 K = g^{\mu\nu} \nabla_\mu n_\nu \ ,
\end{equation}
where $g^{\mu\nu}$ is the complete bulk metric (in contrast to $\tilde{g}^{\mu\nu}$)
and $n^\mu$ represents the outwardly directed, unitary and normal vector to the constant radius surface.
As it is defined to have unitary norm, the explicit expression for the normal 
vector in the ${t,r,x,y}$ coordinate system is given by
\begin{equation}
 n^\mu = (0,1/\sqrt{g_{rr}},0,0)\ .
\end{equation}
Indeed recall that we are considering a surface defined by a constant value for
the radius so that the normal vector is along the radial direction.
Since the extrinsic curvature \eqref{extrins} is given by a covariant divergence,
it can be written in the following fashion%
\footnote{Consider the covariant divergence of a vector $V$ which is given by
\begin{equation}\label{veto}
 \nabla_m V^m = \partial_m V^m + \Gamma^m_{\ ma}V^a\ .
\end{equation}
When the connection symbol has the first two indexes saturated, it admits the following compact rewriting
\begin{equation}\label{logacov}
 \begin{split}
 \Gamma^m_{\ ma} &= \frac{1}{2} g^{mc} \left\{\frac{\partial g_{ac}}{\partial
x^m} + \frac{\partial g_{cm}}{\partial x^a} - \frac{\partial g_{ma}}{\partial
x^c} \right\} 
 =\frac{1}{2} g^{mc} \left\{ \frac{\partial g_{cm}}{\partial x^a} \right\} 
 =\frac{1}{2} \text{tr} \left(\hat{g}^{-1} \partial_a \hat{g} \right) \\
 &=\frac{1}{2} \text{tr} \ \partial_a \ln \hat{g} 
 =\frac{1}{2} \partial_a \text{tr} \ln \hat{g} 
 =\frac{1}{2} \partial_a \ln \text{det} \hat{g} 
 =\partial_a \ln \sqrt{g} \ ,
 \end{split}
\end{equation}
in which $\hat{g}$ is the metric expressed in matrix notation whereas $g$ is
its determinant. 
From the explicit passages given in \eqref{logacov} we got
\begin{equation}
 \Gamma^m_{\ ma} = \frac{1}{\sqrt{g}} \partial_a \sqrt{g}\ .
\end{equation}
Plugging the compact expression into \eqref{veto} we obtain
\begin{equation}
\nabla_m V^m = \partial_m V^m +\frac{1}{\sqrt{g}} (\partial_m \sqrt{g})\, V^m 
             = \frac{1}{\sqrt{g}} \partial_m (\sqrt{g}\, V^m) \ .
\end{equation}}
\begin{equation}\label{exspa}
 K = g^{\mu\nu} \nabla_\mu n_\nu 
   = \frac{1}{\sqrt{-g}}\,\partial_\mu \left( \sqrt-{g}\ n^\mu\right) 
   = \frac{1}{\sqrt{-g}}\,\partial_r \left(\frac{\sqrt{-g}}{\sqrt{g_{rr}}} \right) \ .
\end{equation}
Also the extrinsic curvature needs to be considered in the large radii region,
\begin{equation}
 K = \frac{3}{L} 
 - \frac{3}{r^3}\, L \, g_{tx}^{(0)} g_{tx}^{(1)} 
 - \frac{3}{4 r^3}\, L^5 \epsilon\, g_{tx}^{(0)} g_{tx}^{(0)} + ...\ .
\end{equation}
The intermediate steps are reported explicitly in Appendix \ref{explicit}.
Keeping only the quadratic terms in the fluctuation fields, we obtain
\begin{equation}
 S_{C.T.} = \int dx^3 \left[ 
 4 g_{tx}^{(0)}g_{tx}^{(1)}
 - r^3 g_{tx}^{(0)} g_{tx}^{(0)}
 +\frac{5}{4} \, L^4 \epsilon\, g_{tx}^{(0)} g_{tx}^{(0)} + ...\ ,
 \right]
\end{equation}
and eventually the renormalized, quadratic action for the vector field fluctuations
\begin{equation}\label{osS2}
 S^{(2)}_{\text{on-shell,REN}} = \int dx^3 \left[ 
 \frac{1}{2} \frac{1}{L^2} A_x^{(0)} A_x^{(1)}
 - 3 g_{tx}^{(0)} g_{tx}^{(1)}
 - \frac{1}{2}\, L^4 \epsilon\, g_{tx}^{(0)} g_{tx}^{(0)} + ...\right]\ .
\end{equation}

\subsection{Vector fluctuations of the metric and temperature gradient}
\label{tem_gra}

The temperature corresponds to the inverse period of the analytically 
continued Euclidean time.
Indicating the complex time with $t$ and the temperature with $T$ we have that 
$ \text{Im}(t) \in \left[0,\frac{1}{T}\right)$. Since the original metric 
in the boundary theory is Minkowski, we have $g_{tt}=-1$.
We define a new rescaled time coordinate 
\begin{equation}
 {\mathfrak t} = \frac{t}{T}\ .
\end{equation}
In this way, the ${\mathfrak t}{\mathfrak t}$ metric component becomes temperature dependent
\begin{equation}
 g_{tt} = -1\ \ \  \rightarrow\ \ \ g_{{\mathfrak t}{\mathfrak t}} = - T^2 \ .
\end{equation}
In this framework, we consider at linear order a small fluctuation of the temperature,
\begin{equation}
 g_{{\mathfrak t}{\mathfrak t}} = - (T + \delta T)^2 = - T (T + 2 \delta T) + ...
\end{equation}
We want to ``transfer'' the dependence of $g_{{\mathfrak t}{\mathfrak t}}$ on $\delta T$
to the off-diagonal metric components by means of a diffeomorphism.

The transformation of the metric components under an infinitesimal diffeomorphism is
\begin{equation}
 \begin{split}
 \delta g_{\mu\nu} = \left({\cal L} g \right)_{\mu\nu} 
                   &= g_{\mu \alpha} \xi^\alpha_{\ ,\nu} + g_{\alpha \nu} \xi^{\alpha}_{\ ,\mu} + g_{\mu\nu,\alpha} \xi^\alpha\\
                   &\longrightarrow \partial_\mu \xi_\nu + \partial_\nu \xi_\mu + (\partial_\alpha g_{\mu\nu}) \xi^\alpha
 \end{split}                   
\end{equation}
where in the last passage we have used the metricity of the connection and the fact that
we are considering flat space-time. Choosing an infinitesimal diffeomorphism
corresponding to the vector $\xi = \xi_{\mathfrak t} d{\mathfrak t}$, we have
\begin{equation}\label{flugot}
 \delta g_{{\mathfrak t}{\mathfrak t}} = 2 \partial_{\mathfrak t} \xi_{\mathfrak t} + 2 T (\partial_{\mathfrak t} \delta T) \xi_{\mathfrak t}\ .
\end{equation}
As we want to absorb the dependence of $g_{{\mathfrak t}{\mathfrak t}}$ on $\delta T$, we require 
\begin{equation}\label{abso}
 \delta g_{{\mathfrak t}{\mathfrak t}} = - T \delta T \ .
\end{equation}
Assuming harmonic temporal dependence $e^{- i {\mathfrak w} {\mathfrak t}}$, we combine and solve \eqref{flugot} and \eqref{abso},
obtaining
\begin{equation}
 \xi_{\mathfrak t} = \frac{i}{{\mathfrak w}} T \delta T + ...
\end{equation}
We started with vanishing off-diagonal metric component, but after the diffeomorphism we get
\begin{equation}\label{diffea}
 \delta g_{{\mathfrak t}x} = \frac{i}{{\mathfrak w}} T (\partial_x \delta T) + ...
\end{equation}
where the dots indicate higher powers of $\delta T$.
Eventually we scale back to the original time coordinate
\begin{equation}\label{vai}
 \delta g_{tx} = \frac{i}{\omega}  \frac{\partial_x \delta T}{T} + ...
\end{equation}
We have then related a vector mode fluctuation of the bulk metric to tha gradient of 
a temperature fluctuation.

\subsection{Heat flow and electrical fields}
\label{heaflo}

Consider an infinitesimal reparametrization transformation. 
It produces a variation of the vector potential given by
\begin{equation}\label{gauvara}
 \delta_g A_\mu = A_\nu \xi^\nu_{\ ,\mu} + A_{\mu,\nu} \xi^\nu\ .
\end{equation}
The label $g$ reminds us that the variation of the gauge potential is
due to a reparametrization of the geometry.
If we are initially in Minkowski metric, the covariant derivatives in \eqref{gauvara}
are in fact normal partial derivatives.
Now we specify to the reparametrization transformation induced by the following vector field 
\begin{equation}\label{repanosca}
 \xi_{t} = i \frac{\delta T}{\omega T} \ , \ \ \
 \xi_x = 0 \ .
\end{equation}
Note that this is the same transformation as in \eqref{diffea} but this 
time no temporal rescaling is considered.

Plugging \eqref{repanosca} in \eqref{gauvara} we obtain the gauge potential 
variation induced by the infinitesimal reparametrization along the vector 
\eqref{repanosca}, namely
\begin{equation}
 \delta_g A_i = -A_t \partial_i \xi_t \ .
\end{equation}
So, at the boundary we have
\begin{equation}
 \delta_g A_i = - i \mu \frac{\partial_i \delta T}{\omega T}\ .
\end{equation}

In general, the fluctuation of the gauge potential (i.e. its total variation)
receives two contributions: one is produced by the temperature fluctuation 
through the metric (as just seen), the second is instead related to a proper,
external electrical field variation,
\begin{equation}\label{totavara}
 \begin{split}
  \delta A_i &= \delta_g A_i + \frac{E_i}{i \omega}
  = - i \mu\ \frac{\partial_i \delta T}{\omega T}  + \frac{E_i}{i \omega} 
  = - \mu\ \delta g_{i t}  + \frac{E_i}{i \omega} 
 \end{split}
\end{equation}
The last passage relies on \eqref{vai}.
From \eqref{totavara} we can relate the electrical field to the fluctuations of the bulk fields 
as follows
\begin{equation}
 E_\mu = i \omega \left(\delta A_\mu + \mu\, \delta g_{i t}\right)\ .
\end{equation}
Considering the same set of variations of the bulk fields, we have the 
following variation of the Hamiltonian
\begin{equation}\label{var_ham}
 \delta {\cal H} = \int d^d x \left( T^{\mu\nu} \delta g_{\mu\nu} + J^\mu \delta A_\mu \right)\ .
\end{equation}
Expressing the variation of the fields it in terms of the temperature gradient fluctuation 
(see \eqref{vai}) and the electric field we eventually obtain
\begin{equation}\label{var_ana}
 \delta {\cal H} = \int d^d x \left[ \left(T^{i t} - \mu J^i \right) \frac{\partial_i \delta T}{i \omega T} + J^\mu \frac{E_\mu}{i \omega} \right]\ ,
\end{equation}
from which it arises manifestly that the response to a temperature gradient fluctuation (therefore
what we define as the \emph{heat flow}) corresponds to
\begin{equation}
 Q^i = T^{it} - \mu J^i\ .
\end{equation}

\subsection{The conductivity matrix}

The linear response to perturbations of the generic external source $\phi$
is encoded by the corresponding current $J$. Mathematically, the relation 
between the source perturbation and the induced current is given by the 
retarded Green function  
\begin{equation}\label{Green}
 \delta \langle J^a \rangle =  \langle J^a J^b \rangle\ \delta\phi^b = G^R_{ab}\ \delta\phi^b\ \ .
\end{equation}
So far we do not specify the treatment to a particular kind of field/operator and then 
we have introduced generic indexes $a$ and $b$ to label the sources and the currents.
To account for the presence of the sources $\phi$ we add
the following source/current term in the action of the boundary theory
\begin{equation}
 \sum_a J^a \phi_a \ .
\end{equation}

The Green functions $G_R$ (or retarded correlators) describe the linear response of the system 
to external source perturbations. If the source variation under exam is due to an external electric field, the 
correlator is then related to the electric conductivity of the system.
We know how to calculate the correlators in the holographic framework, so we are able to 
quantitatively describe the linear response of the strongly coupled boundary theory.
In practice,  the correlators can be computed by studying the on-shell action of the gravitational dual system.
As we focus on the linear response we can introduce a simplification to our treatment, 
namely we retain only the linear part of the equations of motion. Equivalently, this 
corresponds to keeping in the action the terms which are at the maximum quadratic in the 
fluctuation fields.
As we have seen in the previous Section, the quadratic on-shell bulk action for the 
fluctuations can be expressed in terms of just boundary contributions.
To this purpose, we need to use the equations of motion for both the background fields and 
the fluctuation fields.

On general grounds, we can express the linear response of the system by means of the 
conductivity matrix, namely
\begin{equation}\label{con_mat}
 \left(\begin{array}{c}
        J_x \\ Q_x
       \end{array}\right) = 
       \left( \begin{array}{cc}
               \sigma & \alpha T\\
               \alpha T & \kappa\, T 
              \end{array}\right)
              \left( \begin{array}{c}
                      E_x \\
                      -(\nabla_x T )/T
                     \end{array}\right)\ ,
\end{equation}
where $J_x$ denotes the electric current and $Q_x$ is the heat flow. 
The diagonal entries of the conductivity matrix are the electric conductivity 
$\sigma$ and the thermal conductivity $\kappa$ (multiplied by the temperature). The off-diagonal 
entries encode mixed transport properties; in this case, thermo-electric effects.
The conductivity matrix is symmetric whenever the equilibrium state is time-reversal 
symmetric this being a consequence of the Onsager argument (see for instance \cite{Musso:2012sn}).

From an asymptotic near-boundary analysis of the fluctuation equations,
we derived the UV behavior for the fluctuation fields given in Equation \eqref{UV_fluct}.
Hence, the solution to the metric fluctuation equation \eqref{metflu}
can be expressed as follows
\begin{equation}\label{sta}
g_{tx} = r^2 \left( g_{tx}^{(0)} + \int_r^\infty dr\, \frac{\phi' A_x }{r^2} \right)\, ,
\end{equation}
from which we have that%
\footnote{In this foot-note we report the explicit passages. We rewrite \eqref{sta}
as 
\begin{equation}\label{pass}
 \frac{1}{r^2} g_{tx} = g_{tx}^{(0)} + \int_r^{+\infty} dr \frac{\phi' A_x}{r^2}\ .
\end{equation}
We then consider the UV expansion of $g_{tx}$
\begin{equation}
 \frac{1}{r^2} g_{tx} = g_{tx}^{(0)} - \frac{1}{r^3} g_{tx}^{(1)} + ...\ .
\end{equation}
Equation \eqref{pass} becomes
\begin{equation}
 - \frac{\phi' A_x}{r^2} = \frac{3}{r^4} g_{tx}^{(1)} + ... \,
\end{equation}
and, expanding the gauge field as well, we obtain
\begin{equation}
 -\frac{1}{r^2} \left( -\frac{1}{r^2} \rho + ... \right)
 \left( A_x^{(0)} + ... \right) = \frac{3}{r^4} g_{tx}^{(1)} + ...\ ,
\end{equation}
from which we eventually have \eqref{guno}.}
\begin{equation}\label{guno}
g_{tx}^{(1)} = \frac{\rho}{3}A_x^{(0)} \, .
\end{equation}

The energy-momentum tensor and the electro-magnetic current
encode the linear response of the system to perturbations
of the metric and the gauge field respectively,
\begin{equation}\label{act_qua}
 \delta  S^{(2)}_{\text{on-shell,REN}} = \int dx^3 \left[ 
 T^{\mu\nu} \delta g^{(0)}_{\mu\nu} + J^\mu \delta A^{(0)}_\mu\right]\ .
\end{equation}
This coincides with \eqref{var_ham}. 
We want to map the variations of the metric and the gauge field to variations 
of physical quantities, namely the temperature gradient and the electro-magnetic 
field. From the argument described in \eqref{tem_gra} we have
\begin{equation}
 \delta g_{tx}^{(0)} = -\frac{i}{\omega} \left(\frac{-\nabla_x \delta T}{T}\right)\ .
\end{equation}
The connection between the vector fluctuations of the metric and the thermal gradient is
therefore straightforward. 

The gauge field, being a vector quantity, is sensitive to
coordinate transformations. In other terms, the gauge field transforms under diffeomorphisms
and, as we have seen, a metric variation induces a gauge field variation. We can split the total gauge 
field variation in a ``pure'' electro-magnetic part and a part induced by the metric variation.
The latter is referred to as ``thermal''. We have
\begin{equation}
 \delta A_x^{(0)} = \delta A_x^{(\text{el})} + \delta A_x^{(\text{therm})}\ ,
\end{equation}
where the purely electro-magnetic variation is due to an electro-magnetic field variation
while the thermal part is proportional to the temperature gradient,
\begin{equation}
 \delta A_x^{(\text{el})} = -\frac{i}{\omega}\, \delta E_x , \ \ \ \
 \delta A_x^{(\text{therm})} = - \mu\, \delta g_{tx}^{(0)} = \mu\, \frac{i}{\omega} \left(\frac{-\nabla_x \delta T}{T}\right)\ .
\end{equation}
Taking stock of the preceding formul\ae~we have
\begin{equation}\label{map_var}
 \delta A_x^{(0)} = -\frac{i}{\omega}\, \left[\delta E_x - \mu\, \left(\frac{-\nabla_x \delta T}{T}\right)\right]\ , \ \ \ \
 \delta g_{tx}^{(0)} = -\frac{i}{\omega} \left(\frac{-\nabla_x \delta T}{T}\right)\ .
\end{equation}
Inserting \eqref{map_var} into \eqref{act_qua} we obtain
\begin{equation}
 \delta  S^{(2)}_{\text{on-shell,REN}} = -\frac{i}{\omega} \int dx^3 \sqrt{-g^{(0)}} \left[ 
 (T^{tx} - \mu\, J^x) \left(\frac{- \nabla_x \delta T}{T}\right) + J^x \delta E_x \right]\ ,
\end{equation}
which is analogous to \eqref{var_ana} used to define the heat transport.

From the map \eqref{map_var} between the variations we get the map between the corresponding derivatives, namely
\begin{eqnarray}\label{map_der}
 \frac{\delta}{\delta E_x} &=& -\frac{i}{\omega} \frac{\delta}{\delta A_x^{(0)}} , \\
 -T\, \frac{\delta}{\delta \nabla_x T} &=& -\frac{i}{\omega} \left[\frac{\delta}{\delta g_{tx}^{(0)}} - \mu \frac{\delta}{\delta A_x^{(0)}}  \right]\ ,
\end{eqnarray}
where we have taken into account that
\begin{equation}
  \delta \nabla_x T = \nabla_x \delta T \ .
\end{equation}
Equations \eqref{map_der} are useful in order to compute correlators by functionally differentiating
the on-shell action.

Let us consider the renormalized on-shell action for the fluctuations up to the quadratic 
order. Take \eqref{osS2} with $L=1$, namely
\begin{equation}\label{actflu}
 S^{(2)}_{\text{on-shell,REN}} = \int dx^3 \left[ 
 \frac{1}{2} A_x^{(0)} A_x^{(1)}
 - 3 g_{tx}^{(0)} g_{tx}^{(1)}
 - \frac{\epsilon}{2}\, g_{tx}^{(0)} g_{tx}^{(0)} + ...\right]\ .
\end{equation}
We first study the derivatives of the fluctuation action \eqref{actflu} with respect to the 
fields $A_x^{(0)}$ and $g_{tx}^{(0)}$, we later translate the results in terms
of the physical quantities. The first derivative with respect to $A_x^{(0)}$ gives
\begin{equation}
 \frac{\delta S^{(2)}}{\delta A_x^{(0)}} = \frac{1}{2} A_x^{(1)} + \frac{1}{2} A_x^{(0)} \frac{\delta A_x^{(1)}}{\delta A_x^{(0)}} 
 - 3\, g_{tx}^{(0)} \frac{\delta g_{tx}^{(1)}}{\delta A_x^{(0)}}\ ,
\end{equation}
where we have regarded the variation of $A_x^{(0)}$ and $g_{tx}^{(0)}$ as independent, so
\begin{equation}
 \frac{\delta g_{tx}^{(0)}}{\delta A_x^{(0)}} = 0\ .
\end{equation}
Similarly, the derivative of the action with respect to $g_{tx}^{(0)}$ is
\begin{equation}
 \frac{\delta S^{(2)}}{\delta g_{tx}^{(0)}} = -3\, g_{tx}^{(1)} - \epsilon\, g_{tx}^{(0)}\ ,
\end{equation}
where again we have recalled the independence of the sources; this time we used
\begin{equation}
 \frac{\delta A_{x}^{(0)}}{\delta g_{tx}^{(0)}} = 0\ .
\end{equation}
Moving forward to the second derivatives, we have
\begin{equation}
 \frac{\delta^2 S^{(2)}}{(\delta A_x^{(0)})^2} = \frac{\delta A_x^{(1)}}{\delta A_x^{(0)}}\ ,
\end{equation}
where we have neglected the second derivatives of the UV subleading terms $A_x^{(1)}$ and
$g_{tx}^{(1)}$, namely
\begin{equation}
 \frac{\delta^2 A_x^{(1)}}{(\delta A_x^{(0)})^2} = 0\ , \ \ \ \
 \frac{\delta^2 g_{tx}^{(1)}}{(\delta A_x^{(0)})^2} = 0\ ;
\end{equation}
this comes from the linearity assumption of the response.
The second derivative with respect to $g_{tx}^{(0)}$ gives
\begin{equation}\label{duvarg}
 \frac{\delta^2 S^{(2)}}{(\delta g_{tx}^{(0)})^2} =  - \epsilon\ ,
\end{equation}
where we have used \eqref{guno} which implies 
\begin{equation}
 \frac{\delta g_{tx}^{(1)}}{\delta g_{tx}^{(0)}} = 0\ .
\end{equation}

Eventually, let us consider the mixed second derivatives 
\begin{equation}\label{GJT}
 \frac{\delta^2 S^{(2)}}{\delta A_x^{(0)} \delta g_{tx}^{(0)}}
 = \frac{\delta^2 S^{(2)}}{\delta g_{tx}^{(0)} \delta A_x^{(0)}}
 = - 3 \frac{\delta g_{tx}^{(1)}}{\delta A_x^{(0)}}\ .
\end{equation}
Notice that we have used
\begin{equation}
 \frac{\delta A_x^{(1)}}{\delta g_{tx}^{(0)}} = 0\ ;
\end{equation}
this comes from the observation that the $A$ fluctuations are governed by
Equation \eqref{traflu} where the metric fluctuations do not appear.


In order to obtain the correlators through differential differentiation 
with respect to the physical quantities $E_x$ and $\nabla_x T$, we recall the map between 
the derivatives \eqref{map_der}, we have
\begin{equation}
 \frac{\delta S^{(2)}}{\delta E_x} = - \frac{i}{\omega}\, \frac{\delta S^{(2)}}{\delta A_x^{(0)}} = - \frac{i}{\omega} A_x^{(1)}\ ,
\end{equation}
\begin{equation}
 -T\, \frac{\delta S^{(2)}}{\delta \nabla_x T} = - \frac{i}{\omega} \left[ - \rho A_{x}^{(0)} -\epsilon g_{tx}^{(0)} - \mu A_x^{(1)} - g_{tx}^{(0)} \rho\right] \ ,
\end{equation}
where we have used \eqref{guno} and the linear assumption
\begin{equation}
 \frac{\delta A_{x}^{(1)}}{\delta A_x^{(0)}} \sim \frac{A_x^{(1)}}{A_x^{(0)}}\ .
\end{equation}

To conclude we give the explicit formula for the electric conductivity
\begin{equation}
 \sigma = i \omega  \frac{\delta^2 S^{(2)}}{(\delta E_x)^2} = - \frac{i}{\omega} \frac{\delta A_x^{(1)}}{\delta A_x^{(0)}}\ .
\end{equation}
Where the factor of $i \omega$ in the definition of $\sigma$ reflects the fact 
the that the electromagnetic current is obtained as the response to a gauge-potential variation.
The other transport coefficients, i.e. the thermal and thermo-electric conductivities 
introduced in \eqref{con_mat}, can be similarly and precisely defined%
\footnote{A neat treatment of the thermal and thermo-electric response of the s-wave superconductor
is postponed to \cite{progresso} where particular attention is paid to the question of counter-terms.
A related and interesting discussion about counter-terms in holography is given in \cite{Argurio:2013uba}.}.


\subsection{Numerical results for the conductivity}

Consistently with the ansatz considered insofar, we have to solve systems of ordinary 
differential equations both to study the background and the fluctuations around it.
Nevertheless these are already rather complicated mathematical problems and we need to 
resort to numerical method in order to solve them. Before entering into any specific detail of the numerical
procedure we briefly illustrate the results about the linear response of the holographic superconductor.

\begin{figure}[ht]
\centering
\includegraphics[width=70mm]{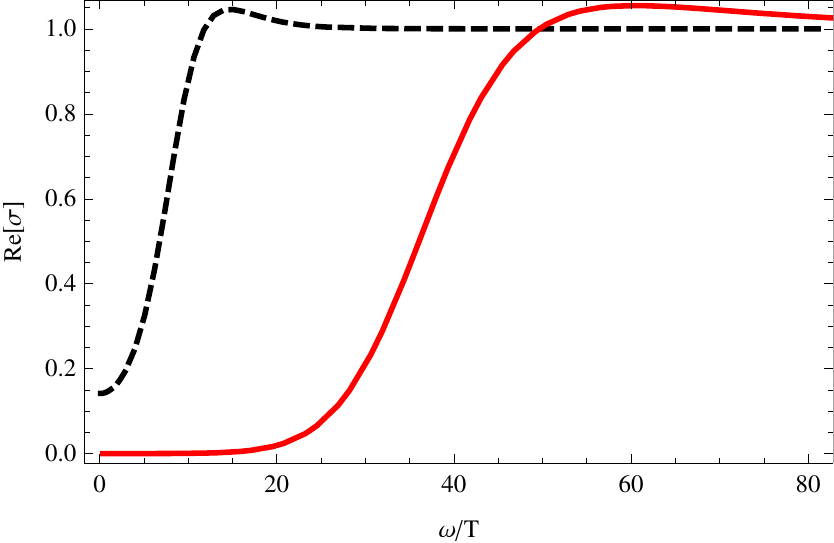} 
\includegraphics[width=70mm]{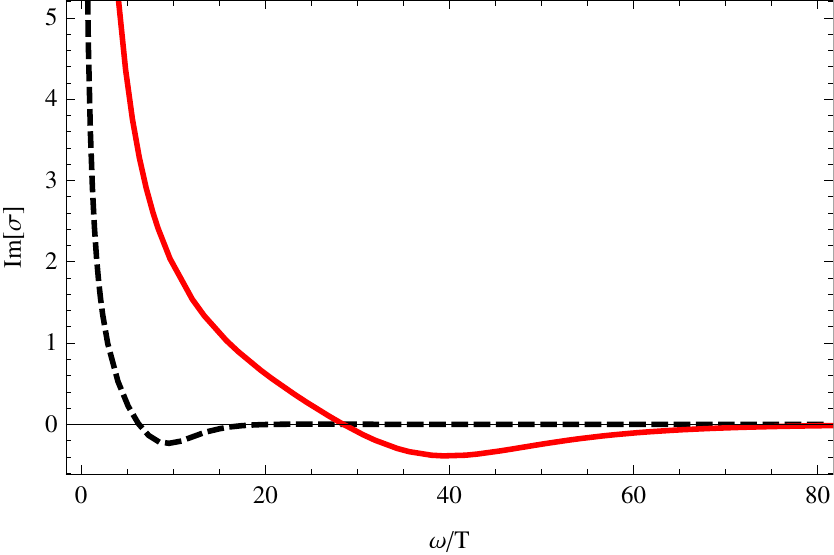} 
\caption{Real (left) and imaginary (right) part of the electric conductivity. The dashed lines refer 
to criticality, i.e. $T=T_c$. The solid lines correspond instead to $T\sim 0.3\, T_c$.}
\label{ele_condu_fig}
\end{figure}

In Figure \ref{ele_condu_fig} the real and imaginary parts of the electrical conductivity are reported 
as functions of the frequency of the electric source perturbation. At small values of the frequency 
we observe the presence of a ``gap-like'' depletion region where the real part of the conductivity is suppressed.
The feature is apparent already at the critical temperature while it disappears at very high temperature (with 
respect to the chemical potential). We are already alluding to the fact that such a depletion region is not 
automatically a hallmark for superconductivity. In spite of the similar qualitative behavior at $T_c$ and 
at lower temperature, we observe that in the latter case the ``bottom'' of the depletion region approaches a 
hard gap, meaning that it seems to attain a vanishing value. More comments on all these features are given below.

At higher frequency, the real part of the electric conductivity tends to a constant whose unitary value 
can be expected from self-duality arguments of the bulk $4$-dimensional theory, see \cite{Herzog:2007ij} for further detail.
The high-frequency behavior is universal and shared by both the normal and the condensed phase. 
Indeed whenever the probing frequency is much above the typical scales of the systems we do not distinguish between 
the phases.

The imaginary part of the electrical conductivity gives us important information at low-frequency.
Actually we observe the occurrence of a pole at vanishing $\omega$. From causality arguments, one 
can derive the so-called Kramers-Kronig relations which map the presence of a pole in the imaginary part 
of the conductivity to a delta function in the real part and at the same position of the pole.
Therefore, the presence of a pole in the imaginary part is signaling the presence of a diverging real 
part of the conductivity at null frequency. Again we have a delta function both at $T_c$  and in the 
superconducting phase. The spectral weight of the delta function corresponds to the missing spectral weight 
of the low-frequency depletion region described above.

A puzzle is served: it seems to have a system who is already superconducting at $T_c$. To solve the question 
we have to recall that the system under study is translationally invariant and charged. It is than natural 
to have a diverging conductivity also in the normal phase. Such infinite conductivity for $\omega = 0$ 
however is not the ``authentic'' superconductivity contributed by the scalar condensate. We will shortly 
and quantitatively study this latter in the next Subsection.


%

\subsection{Authentic superconductivity}
\label{autsuc}

As we have already mentioned, translation invariance for a charged system leads 
to a diverging conductivity at zero frequency. The divergence being related to the 
constant acceleration driven by the electric source against which no resistance is 
provided by momentum dissipating scattering. To claim that a translationally invariant 
and charged system is actually a superconductor one has to disentangle the two contribution
to the D.C. diverging conductivity. 

One can proceed studying the amplitude of the delta function at $\omega = 0$ of the 
real part of the electrical conductivity. This quantity represents the spectral weight
of the delta itself and ``counts'' the degrees of freedom contributing to the zero-frequency 
electrical transport. A delta function escapes a direct numerical observation, nevertheless we have
quantitative information studying the residue of the corresponding pole in the imaginary part.
Indeed this is mapped to the sought for area by the Kramers-Kronig relations.

\begin{figure}[ht]
\centering
\includegraphics[width=100mm]{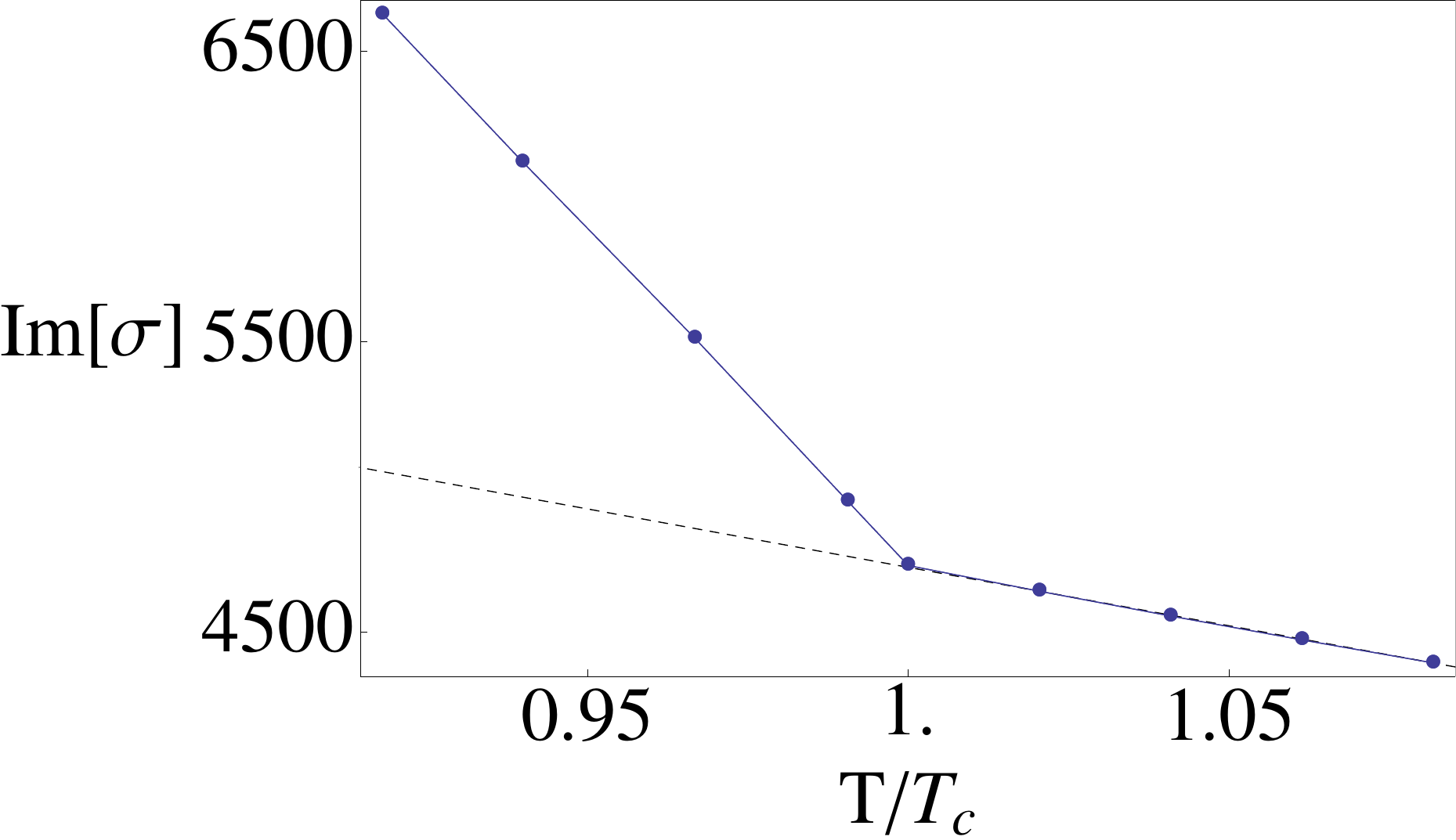} 
\caption{The plot shows the value of the imaginary part of the electric conductivity taken at
a reference frequency well within the depletion region, specifically $\omega^* \sim 5 \cdot 10^{-5}$. 
This quantity is related to the area ``below'' the delta of the real D.C. diverging conductivity. 
We observe a discontinuous (at the level of the first derivative) 
behavior at the transition signaling a new contribution below $T_c$. 
The novel contribution in the condensed phase corresponds to the authentic superconducting behavior.}
\label{AU_SUCO}
\end{figure}

In practice we choose a reference value for the frequency, say $\omega^*$, which has to be very small 
with respect to the characteristic size of the depletion region. In other terms, our 
reference frequency must fall well within the depletion region itself and be in this sense close to the origin.
We then look at the value of Im$[\sigma(\omega^*)]$ at the reference frequency and vary the temperature.
We do not need to perform a fit to actually measure the residue, what we simply need is to characterize a 
discontinuous behavior at the phase transition, i.e. when lowering $T$ we pass through $T_c$. The data collected 
are summarized in Figure \ref{AU_SUCO} where a discontinuity at the level of the first derivative is apparent.
This signals a novel contribution below $T_c$ which increases the intensity of the D.C. diverging real superconductivity.

As a final comment it is interesting to mention that a similar study has been performed also on the 
unbalanced holographic superconductor \cite{Bigazzi:2011ak} where a superconducting phenomenon has been 
reported also in the ``spin-spin'' channel. This spin superconductivity is found through an analysis which 
follows the same lines as that just described. The phenomenon is there particularly interesting because the 
spin superconductivity seems to arise because of a complicated intertwined spin-electro dynamics and not 
from a spontaneous breaking of the spin symmetry (accounted for effectively by a second $U(1)$ bulk gauge field).

\subsection{Thermodynamic and spectroscopic gaps}
\label{carrier}

In a superconductor there are two definitions for the gap:
the \emph{spectroscopic} and the \emph{thermodynamic} one.
The former is related to the gap $\omega_g$ showed by the optical conductivity;
the frequency $\omega_g$ corresponds to the energy that an external incident 
wave must have in order to be ``tuned'' with the Cooper pairs and then break them
splitting the two paired electrons. The thermodynamic gap $\Delta$ is instead related 
to the energy of a single quasi-particle excitation. We can measure it
observing the real part of the optical conductivity at very low frequency and, 
in particular, its temperature dependence. Indeed we have
\begin{equation}
 \lim_{\omega\rightarrow 0 } \text{Re}(\sigma) \sim e^{-\Delta/T}\ .
\end{equation}

In a superconductor where the superconducting mechanism is based on a pairwise interaction,
like in the BCS theory, the two gaps are related by
\begin{equation}\label{gaps}
 \omega_g = 2 \Delta\ .
\end{equation}
The factor of two is due two the fact that $2$ electrons participate to the pairing mechanism.
Understanding the origin of \eqref{gaps} is slightly tricky as one could guess that the two 
gaps had to be equal. Actually, the thermodynamic gap is sensitive to the energy needed to 
the individual quasi-particle to be excited, while the spectroscopic gap is associated to the breaking 
of a pair. Of course exciting one single electron out of a pair means breaking the pair itself.
However we have to think about a large number of contemporary events and not focus on the single 
pair. Said otherwise, when the energy fluctuations allow some electrons to leave their pairs, 
the unpaired electrons can rearrange to form new pairs without needing to be excited as well. 

The study of the relation between the two energy gaps contains information about the 
superconducting mechanism. It is then interesting two analyze what happens in the holographic
superconductor. Numerical analysis carried on in \cite{Hartnoll:2008vx} showed that for the 
holographic superconductor we have
\begin{equation}
 \omega_g = \alpha\, \Delta\ ,
\end{equation}
where $\alpha$ is not necessarily equal to $2$. Actually, $\alpha$ depends on the 
details of the model and the particular values of its parameters. We then understand that 
the holographic superconductor (which, we remind the reader, is a bottom-up model) appears
not to predict a specific pairwise (nor $n$-uple) mechanism underlying superconductivity 
at strong coupling.

In a top-down approach 
it could be fixed by the UV completion. 

\subsection{No hard gap at $T=0$}

A hard gap signals the lack of modes in the frequency region $\omega<\omega_g$.
 Indeed, ``hard gap'' refers to the exact vanishing of the real part of the optical conductivity
 in the window $0<\omega<\omega_g$ at $T=0$.
 This contrasts with the presence of Goldstone (and multi-Goldstone) bosons in the 
 spontaneously broken phase of the superconductor. Actually, it has been shown in 
 \cite{Horowitz:2009ij} that the holographic superconductor has no hard gap.  Nonetheless,
 a precise understanding about the origin of the low frequency spectral weight and its
 possible relation with Goldstone bosons is still a question to be clarified.
 For instance, it is not yet clear whether such contribution from Goldstone bosons 
 remains finite in the large $N$ limit.
  
 The optical conductivity can be expressed in terms of a reflection coefficient in an
appropriate one-dimensional Schr\"{o}dinger problem obtained by means of a change of coordinate
in the radial equation for $A_x$. In particular, we have an expression of the following form
\begin{equation}
 \sigma(\omega) = \frac{1-{\cal R}}{1+{\cal R}}\ ,
\end{equation}
where ${\cal R}$ represents a reflection coefficient in the Schr\"{o}dinger scattering problem.
This claim has a general valence and it holds true whenever the quadratic action for the 
Maxwell field has the standard form. For explicit details we refer to \cite{Horowitz:2009ij}.
This map translating the computation of the optical conductivity into a $1$-dimensional
scattering problem on a potential has been the key technical step to prove the absence of a hard 
gap in the holographic superconductor.

\section{Comments on the numerical approach and the shooting method}
\label{shoot}

The actual solution of the dual system of ordinary differential equations describing the 
equilibrium and the transport properties of the boundary medium needs the so-called shooting method.
Actually, as argued in the preceding sections, either for consistency reasons or to fix the external physical parameters 
we need in general to impose boundary conditions both at the horizon and at the conformal boundary.
It is just this necessity of fixing conditions at different points which calls for the shooting method
which corresponds to the complete fixing of all the boundary condition at a single point
to values that are then tuned to satisfy other requirements elsewhere.
Actually the shooting method translates a boundary value problem
to an initial value problem. Let us try to clarify the point with an example.

Consider a boundary value problem of a second-order ordinary differential equation where 
the unknown function is $y$ while the independent variable is $t$. Take
\begin{equation}\label{bou_pro}
 y''(t) = f[t,y(t),y'(t)]\ , \ \ \ \ y(t_0) = y_0 \ , \ \ \ \ y(t_1) = y_1\ ,
\end{equation}
as the definition of the boundary value problem we want to solve.
We then introduce a parametric dependence and define $y(t;a)$ that indicates the solution of the 
initial value problem
\begin{equation}
 y''(t) = f[t,y(t),y'(t)]\ , \ \ \ \ y(t_0) = y_0 \ , \ \ \ \ y'(t_0) = a\ ,
\end{equation}
where $a$ represents the initial velocity.
We can therefore define the function $F(a)$ as the difference between $y(t_1;a)$ and the particular boundary value $y_1$
that we have introduced in \eqref{bou_pro},
\begin{equation}
 F(a) = y(t_1;a) - y_1
\end{equation}
If the boundary value problem \eqref{bou_pro} possesses a solution, then $F$ presents a root.
Furthermore such root corresponds to the value of $y'(t_0)$ that leads to a solution $y(t)$ of the boundary problem
specified in \eqref{bou_pro}.

\section{Unbalanced holographic superconductors and multiple orderings}

An important motivation for studying the occurrence of multiple orderings
in a strongly coupled context concerns the physics of high-$T_c$ superconductors.
Such systems actually feature a particularly complicated phase diagram where 
different order parameters coexist and mutually influence one another. 
Striped phases and charge density waves, anti-ferromagnetism and spin density waves
are among the possibilities one can encounter in the study of high-$T_c$ superconductivity%
\footnote{To have a wider though succinct review see \cite{Zaanen:2010yk}.}.
A better understanding of the interlaced dynamics of the different orderings 
could shed light on the physics of the so-called pseudo-gap phase and even the 
mechanism of high $T_c$ superconductivity itself. 

One family of systems one can employ to have holographic models of coexisting orders
is inspired by the unbalanced holographic superconductor. This latter is essentially a generalization 
of the standard holographic superconductor with the addition of a second Abelian gauge field in the bulk.
Along the same lines described above, this second gauge fields describes from a dual perspective 
the presence of a second chemical potential associated to a second type of conserved charge.
As argued in \cite{Bigazzi:2011ak} the unbalanced holographic superconductor can be interpreted as 
a superconductor where the population of spin-up and spin-down electrons have different chemical potential
and is hence unbalanced.

Among the simplest holographic models featuring two order parameters is the further addition to the unbalanced
holographic superconductor of a second scalar field (charged under the ``second'' gauge field). 
The resulting system resembles a double copy of the 
standard holographic superconductor. However either a direct coupling between the scalars or backreaction
(or also both at a time) make the interlaced dynamics of the two orderings richer.
In particular one can address in a simple and controlled system questions like the coexistence and competition 
of different orderings \cite{Musso:2013ija}.

\begin{figure}[ht]
\centering
\includegraphics[width=90mm]{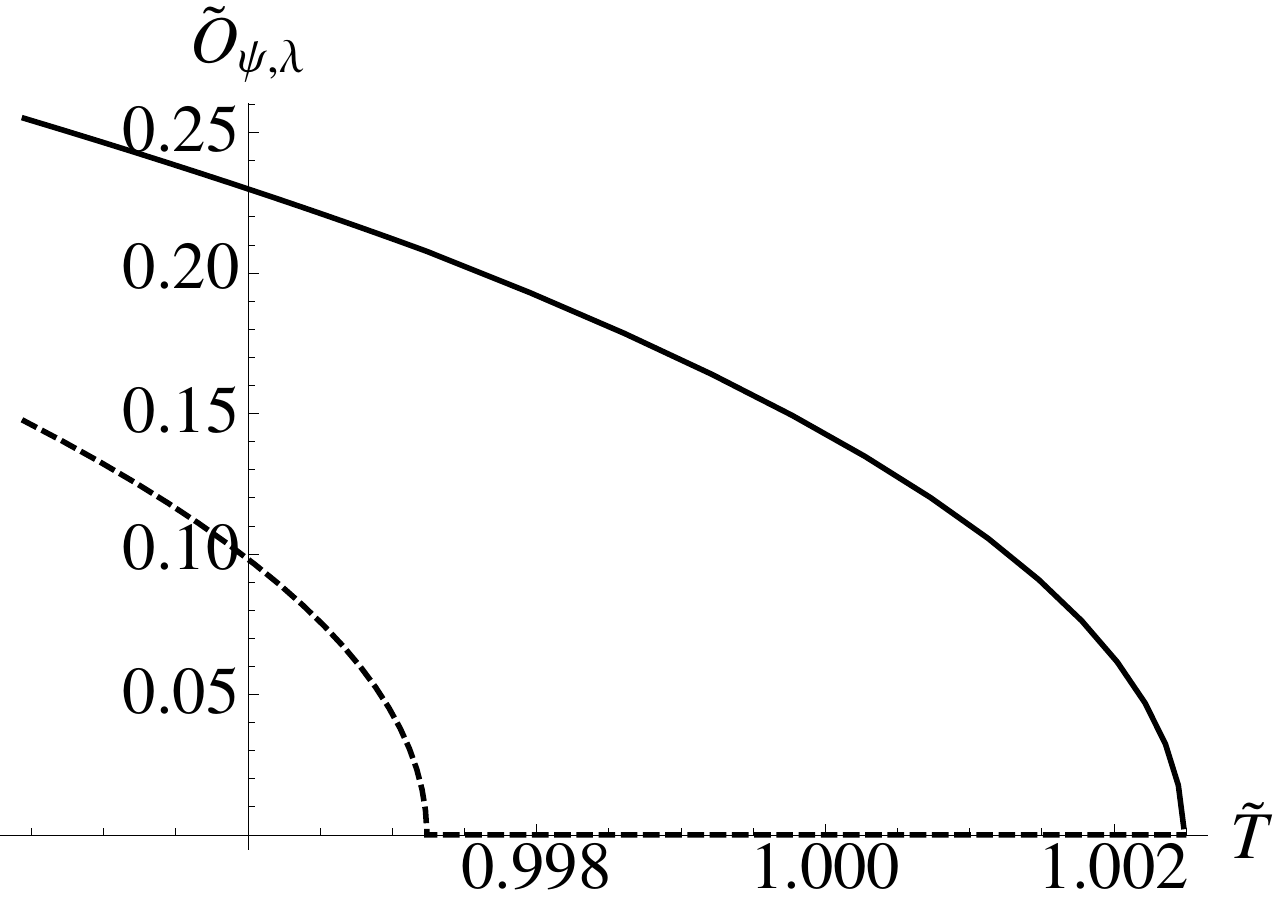} 
\label{xiste}
\caption{Plot showing the possibility of having coexistence of two scalar condensates at the same time.}
\end{figure}

\begin{figure}[ht]
\centering
\includegraphics[width=70mm]{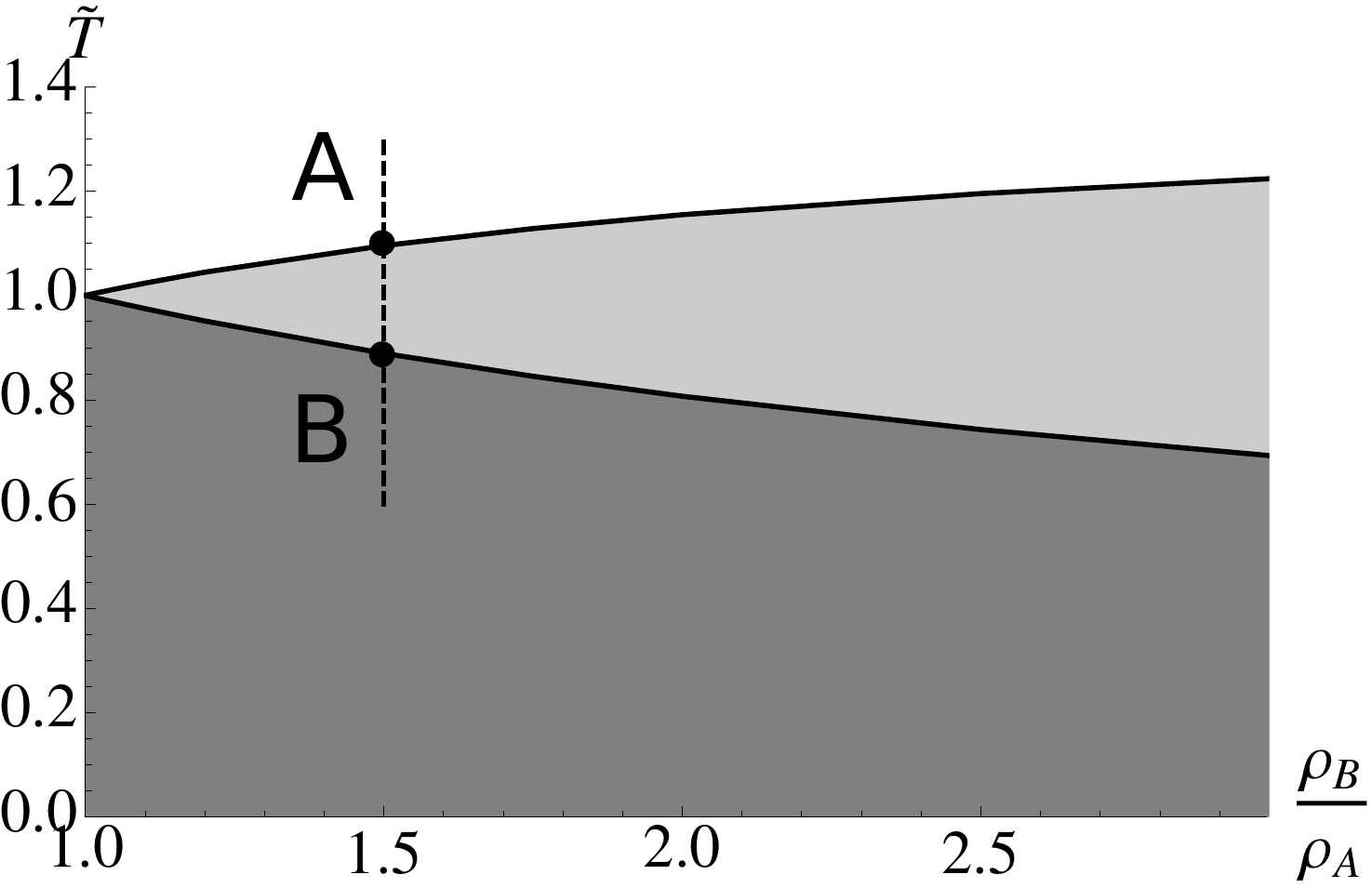} 
\includegraphics[width=70mm]{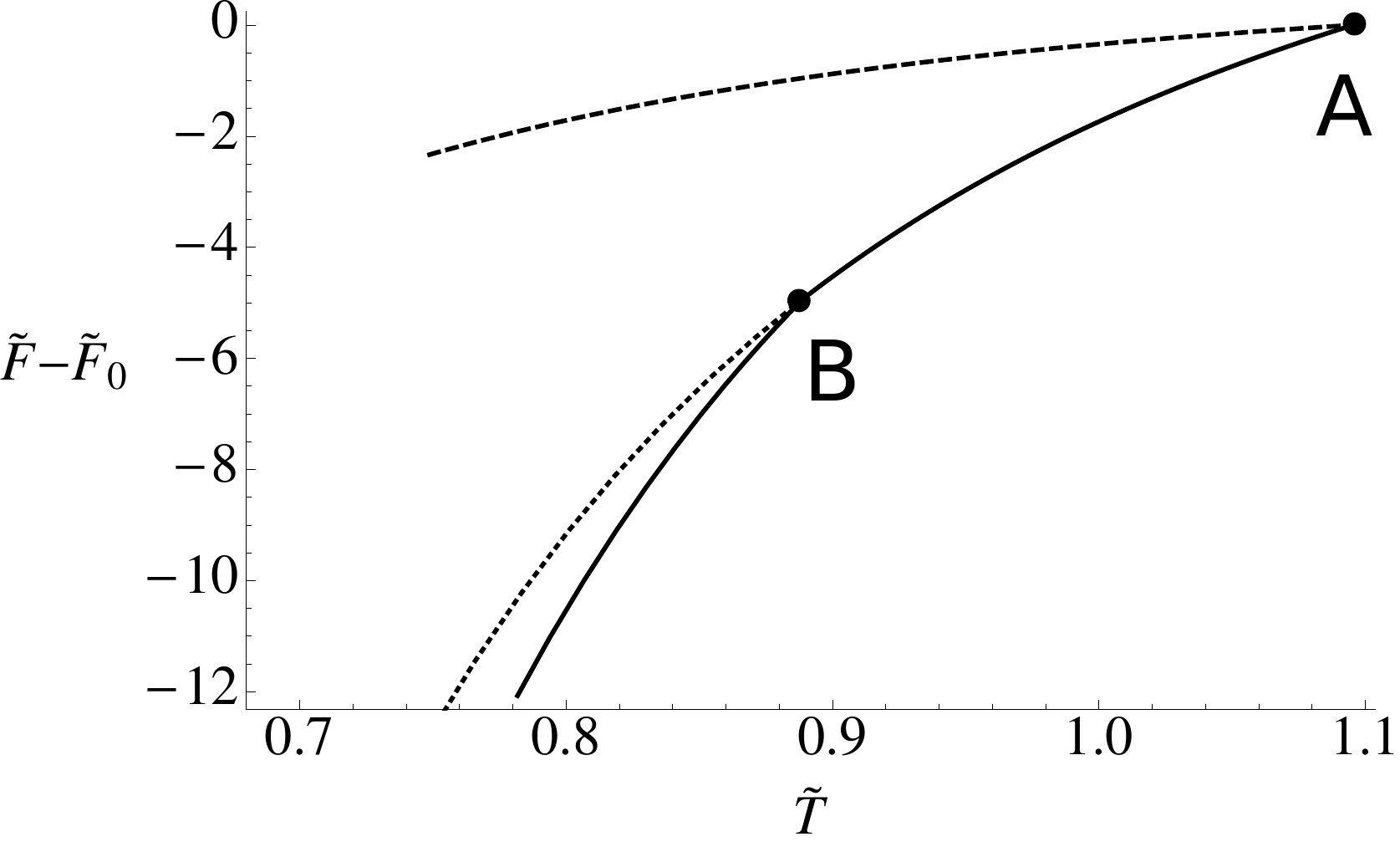} 
\label{eq_sb}
\caption{Portion of the phase diagram of the unbalanced holographic superconductor with two scalars 
described in the main text (left).
The white region corresponds to the normal phase, the pale gray region has a single condensate
whereas the darker gray region corresponds to double condensation of both scalar order parameters.
The solid lines separating the above-mentioned phases correspond to second order transitions.
This is made evident in the free energy diagram reported on the right. The transition features a discontinuity
at the level of the second derivative in the temperature. Note that the energy plot is useful to understand
which phase is the thermodynamically favored one at various values of the temperature.}
\end{figure}

Figure \ref{xiste} is a plot showing the two condensations of the two scalar order parameters.
The first observation is that coexistence of such orderings does occur. The holographic set-up allows 
a more systematic study of the equilibrium. In Figure \eqref{eq_sb} we report 
a phase diagram (left) resulting from a detailed study of the free-energy (right). This latter is 
necessary to distinguish which phase is thermodynamically favored.

Also the standard p-wave holographic superconductor \cite{Gubser:2008wv} allows for a generalization
consisting in a ``doubling'' that is a system with two interacting bulk gauge fields. In \cite{Amoretti:2013oia}
such generalized p-wave has been throughly studied in a probe approximation and an interesting phenomenological 
interpretation had been advanced, that of a ferro-magnetic superconductor. Such systems feature the coexistence 
of multiple orderings, a superconducting and a ferro-magnetic one, whose strongly-coupled dynamics could 
furnish a holographic toy model of a superconductor transition occurring on the border of ferro-magnetism \cite{2007PhRvL,Saxena:2000,Huy:2007}.
In this sense giving a holographic version of a mechanism analogous to that proposed in \cite{Fay:1980wv}.

\section{Some other generalizations and related research}

Throughout these notes we have been mainly concerned with s-wave superconductors
where the order parameter is isotropic and behaves as a scalar with respect to spatial rotations.
However, one can study holographically p and d-wave superconductors as well where the condensing order parameters have
respectively vector (i.e. spin 1) and tensor (i.e. spin 2) properties. The p-wave has been introduced 
in \cite{Gubser:2008wv} and has later been extensively studied; see for instance \cite{Ammon:2009xh,Arias:2012py}
to find an analysis of the backreacted case. The d-wave superconductor is phenomenologically interesting as many real-world
high-$T_c$ superconductors have been shown to posses an order parameter with such a symmetry. From a holographic 
perspective, d-wave superconductors (which require the introduction 
of a charged and massive spin 2 particle in the gravitational bulk) can pose some consistency issue related to the presence
of ghosts, loss of hyperbolicity and faster than light propagation.
Nevertheless, at an effective field theory level, models have been 
introduced and studied, see for instance \cite{Chen:2010mk,Benini:2010pr} and references therein.

One important family of generalized models for holographic superconductors involves the 
introduction of momentum dissipation. As we have discussed before, the translational invariance 
of the standard holographic superconductor yields a diverging contribution to the 
DC conductivity from which one must disentangle the authentic superconductivity.
Translation invariance is also an usually undesirable feature in relation to any phenomenological interpretation.
Actually, finite size and the presence of a lattice are instances that broke such an invariance.

The introduction of a lattice is particularly relevant for condensed matter applications and indeed in recent times
much effort has been spent to describe (or at least mimic) an ion lattice holographically \cite{Horowitz:2012ky,Horowitz:2012gs,Balasubramanian:2013yqa}.
This can be done in a direct way by means of spatially modulated sources (e.g. a spatially inhomogeneous chemical potential).
The task can be technically demanding in that the system of equations usually turns to have partial differential equations 
instead of ordinary ones (dependence on space has to be included).

According to recent developments, instead of considering explicitly spatially modulated fields, one can break 
momentum conservation on the boundary within models featuring explicit diffeomorphism breaking in the bulk.
A viable approach is to consider massive gravity in the bulk \cite{Vegh:2013sk,Andrade:2013gsa,Davison:2013jba,Blake:2013bqa,Blake:2013owa}%
\footnote{The massive gravity framework can be interpreted also in relation to the presence of 
disorder which breaks momentum conservation \cite{Davison:2013txa}.}. Note that the introduction of 
a mass term for the bulk graviton breaks momentum conservation in the boundary theory in a homogeneous 
way.

The massive gravity framework can pose some theoretical difficulty related 
to instability issues and the presence of ghosts and super-luminal propagation.
In spite of this theoretical problems, massive gravity has been checked to yield sensible
holographic results \cite{Blake:2013bqa} and in particular the thermodynamics of massive gravity models 
is sound. The test has shown that the area/entropy law and the first law of thermodynamics hold 
for black holes in massive gravity too and, on top of that, the holographic renormalization 
procedure can be performed as in the massless case without the introduction of new terms that affect the 
thermodynamics (the new finite terms in the bulk on-shell action depend only on the mass terms for the bulk graviton).
In addition, there is a very relevant 
result from holographic massive gravity models: a linear in $T$ resistivity \cite{Blake:2013bqa}.
As we argued at the beginning of these notes, such behavior is the characteristic imprint of 
strange metal behavior. The fact that massive gravity poses consistency issues could be in some sense circumvented;
indeed, taking just inspiration from the holographic models one can argue on general grounds that a 
linear in $T$ resistivity is a consequence of just two assumptions: the presence of quenched (i.e. not dynamical)
disorder and strongly coupled critical liquid behavior (with minimal viscosity $\eta/s \sim 1/4\pi$) \cite{Davison:2013txa}.

Spatial modulation are relevant also in
the perspective of describing the so-called striped phases of high-$T_c$ superconductors. 
In real-world systems such phases are related to the formation of charge density waves and current density
waves and their interplay. A holographic approach to study striped phases and the spontaneous arising of 
spatial modulation we refer to \cite{Donos:2011bh}.

As the holographic approach offers the possibility of treating and analyzing 2-point correlations
and the linear response, two general phenomenological areas which can be naturally explored in the gauge/gravity context are the hydrodynamic behavior
(e.g. sound waves, quasi-normal-modes and on) and light-waves propagation through the holographic medium. 
Regarding the latter, a careful analysis can unveil exotic optical properties like for instance negative refraction 
\cite{Amariti:2010jw,Amariti:2011dj} or additional light-waves \cite{Amariti:2010hw}. The holographic superconductor, 
being a holographic model, is no exception and both its hydrodynamic \cite{Amado:2009ts} and light-propagation 
properties \cite{Amariti:2011dm} can and have been widely studied in the literature.

Another recent and interesting line of research regards the study of ``dirty'' superconductors
where noise or disorder are considered \cite{Arean:2013mta,Zeng:2013yoa}. The holographic approach allows us 
to gain insight on the relation between strong coupling and disorder. Indeed, the question relating 
the interplay between strong interactions (especially the electron-electron interaction) and disorder
(e.g. strong localization) is not yet settled and holographic methods could complement standard approaches.

These notes are concerned with the condensed matter application of some holographic models
describing symmetry breaking at strong coupling. It must be mentioned that similar models are 
applicable to study other phenomenological areas, in particular strongly correlated plasmas and QCD%
\footnote{See for instance \cite{CasalderreySolana:2011us,Bigazzi:2009tc,Bigazzi:2009bk,Bigazzi:2011it,Gursoy:2010fj} and references therein.
For recent studies regarding the transport properties of models dual to D-brane systems we refer to \cite{Tarrio:2013tta} and references therein.}.
By means of holography one can for instance describe color superconductivity \cite{Basu:2011yg} 
and flavor superconductivity \cite{Ammon:2008fc,Ammon:2009fe}.

This final brief panoramic account of the research lines which are related to the basic holographic 
superconductor is of course partial and biased by the author viewpoint. 

%

\section*{Acknowledgments}

A particular acknowledgment goes to Pierre-Henry Lambert for the organization of the IX Modave school and to
Andrea Amoretti who corrected some mistakes of the draft.

I want also to thank Riccardo Argurio, Daniel Arean Fraga, Andrea Mezzalira, Diego Redigolo, Davide Forcella, 
Carlo Maria Becchi, Massimo D'Elia, Ignacio Salazar Landea, John McGreevy, John Bhaseen, Sean Hartnoll, Richard Davison,
Alessandro Braggio, Nicola Maggiore, Nicodemo Magnoli, Hampus Linander, Eduardo Conde Pena, 
Gustavo Lucena G\'omez, Micha Moskovic, Lorenzo Di Pietro, Blaza Oblak, Teresa Bautista Solans, Antonio Amariti, 
Manuela Kulaxizi, Andrea Campoleoni, John Estes, Laura Donnay, Bert Van Pol and Rubens Montens for observations, questions and, 
in general, important and interesting discussions.

\appendix

\section{Explicit passages}
\label{explicit}

Here are reported some expansion useful in the explicit holographic renormalization 
procedure described in the main text. A near boundary limit is understood. Furthermore
only up to quadratic terms in the fluctuations are retained, in line with the linear 
response framework of the main text.
\begin{equation}
 \sqrt{-g} = r^2
 + \frac{1}{2} r^2 L^2 g_{tx}^{(0)} g_{tx}^{(0)}
 +\frac{1}{r} L^2 g_{tx}^{(0)} g_{tx}^{(1)}
 + \frac{\epsilon L^6}{4 r} + ...
\end{equation}

\begin{equation}
 \frac{1}{\sqrt{-g}} = \frac{1}{r^2}
 - \frac{L^2}{2 r^2} g_{tx}^{(0)} g_{tx}^{(0)}
 - \frac{L^2}{r^5} g_{tx}^{(0)} g_{tx}^{(1)}
 - \frac{\epsilon L^6}{4 r^5} g_{tx}^{(0)} g_{tx}^{(0)} + ...
\end{equation}

\begin{equation}
 \frac{1}{\sqrt{g_{rr}}} = \frac{r}{L} - \frac{\epsilon L^3}{4 r^2} + ...
\end{equation}

\begin{equation}
 \begin{split}
 \sqrt{\frac{-g}{g_{rr}}} = & \frac{r^3}{L}
 + \frac{r^3 L}{2} g_{tx}^{(0)} g_{tx}^{(0)}
 + L g_{tx}^{(0)} g_{tx}^{(1)}
 + \frac{\epsilon L^5}{4} g_{tx}^{(0)} g_{tx}^{(0)}
 - \frac{\epsilon L^3}{4}\\
 &- \frac{\epsilon L^5}{8} g_{tx}^{(0)} g_{tx}^{(0)}
 - \frac{\epsilon L^5}{4 r^3} g_{tx}^{(0)} g_{tx}^{(1)}
 - \frac{\epsilon^2 L^9}{16 r^3} g_{tx}^{(0)} g_{tx}^{(0)} + ...
 \end{split}
\end{equation}

\begin{equation}
  \partial_r \sqrt{\frac{-g}{g_{rr}}} =  \frac{3}{L} r^2
  + \frac{3}{2} L r^2 g_{tx}^{(0)} g_{tx}^{(0)}
  + \frac{3}{4} \frac{\epsilon L^5}{r^4} g_{tx}^{(0)} g_{tx}^{(1)}
  + \frac{3}{16} \frac{\epsilon^2 L^9}{r^4} g_{tx}^{(0)} g_{tx}^{(0)} + ...
\end{equation}

\begin{equation}
 \frac{1}{\sqrt{-g}}\,  \partial_r \sqrt{\frac{-g}{g_{rr}}} = 
 \frac{3}{L}
 - \frac{3 L}{r^3} g_{tx}^{(0)} g_{tx}^{(1)}
 - \frac{3}{4} \frac{\epsilon L^5}{r^3} g_{tx}^{(0)} g_{tx}^{(0)} + ...
\end{equation}

\end{document}